# Nuclear temperatures from the evaporation fragment spectra and observed anomalies

A. Ray[1], A. De[2], A. Chatterjee[3], S. Kailas[3], S. R. Banerjee[1], K. Banerjee[1], S. Saha[4]

[1]*Variable Energy Cyclotron Center, 1/AF, Bidhannagar, Kolkata – 700064, India*

[2]*Raniganj Girls' College, Raniganj, West Bengal, India*

[3]*Nuclear Physics Division, Bhabha Atomic Research Center, Mumbai, India*

[4]*Saha Institute of Nuclear Physics, 1/AF, Bidhannagar, Kolkata-700064, India*

**Abstract.** The extreme back-angle evaporation spectra of alpha, lithium, beryllium, boron and carbon from different compound nuclei near A $\approx$ 100 ($E_X$ = 76 – 210 MeV) have been compared with the predictions of standard statistical model codes such as 'CASCADE' and 'GEMINI'. It was found that the shapes of the alpha spectra agree well with the predictions of the statistical models. However the spectra of lithium, beryllium, boron and carbon show significantly gentler slopes implying higher temperature of the residual nuclei, even though the spectra satisfy all other empirical criteria of statistical emissions. The observed slope anomaly was found to be largest for lithium and decreases at higher excitation energy. These results could not be understood by adjusting the parameters of the statistical models or from reaction dynamics and might require examining the statistical model from a quantum mechanical perspective.

## I. INTRODUCTION

The temperature of a system is defined when it is in thermal statistical equilibrium with, in principle, infinite lifetime. The atomic nucleus is a microscopic system and the temperature of a nucleus is determined from the emission of the small parts of the nucleus itself, assuming the formation of long-lived unstable dinuclear states of the fragment and the residual nuclei in the exit channel and the statistical decay of those states. Generally, the temperature of the residual nucleus is determined from the slope of the evaporation neutron, proton or alpha spectra. Charity et al. [1] studied extensively the spectral shapes of the evaporation neutron, proton and alpha particles from a large number of compound nuclei and fitted them using the statistical model code 'GEMINI'. They [1] needed an excitation energy dependent level-density parameter to explain the spectral shapes of the evaporation neutron, proton and alpha spectra from the heavier compound nuclei (A > 150). However in order to understand the evaporation spectra from the lighter compound nuclei (A ≤100), no such excitation energy dependent level-density parameter was required [1].

So far, there has not been any significant effort to understand the spectral shapes of heavier evaporation fragments such as lithium, beryllium, boron and carbon by comparing them with the standard statistical model codes. The absolute cross-sections of such fragments as calculated from the statistical model codes are rather sensitive to many parameters of the statistical model such as the transmission coefficients, critical angular momentum, diffusivity of the spin distribution and the level-density parameter. On the other hand, the spectral shape is largely insensitive to most of those parameters except the level-density parameter. In the case of compound nuclei in the A ≈100 mass region, level-density parameter becomes almost independent of the excitation energy as found by Charity et al. [1]. So it would be somewhat simpler to study and compare the evaporation fragment spectra from the compound nuclei in the A ≈100 mass region with the statistical model codes. In this paper, we have compared the extreme back-angle evaporation spectra of alpha, lithium, beryllium, boron and carbon particles from $^{16}O+^{89}Y$, $^{16}O+^{93}Nb$ and $^{3}He+Ag$ reactions producing the compound nuclei in the A ≈100 mass region (having excitation energies ranging from $E_X$=76 to 210 MeV) with the statistical model codes 'CASCADE' [2] and 'GEMINI' [1]. The statistical character of the spectra has been demonstrated from the observed back-angle rise of the angular distribution and lack of any entrance channel dependence of the spectra for $^{16}O+^{89}Y$ and $^{12}C+^{93}Nb$ reactions forming the same compound nucleus with similar spin distribution and excitation energy. In the case of the $^{3}He+Ag$ reaction, the data was taken from ref [3,4]and the authors concluded from their detailed analysis that the back-angle heavy-ion spectra should be statistical. It has been found from our comparison that the observed spectral shapes of the alpha particles agree well with the statistical model calculations, but the experimental spectra of the heavier fragments (particularly lithium, beryllium and boron) show significantly gentler slope (than the statistical model calculations) implying higher than the expected temperature of the residual nuclei for $^{16}O$, $^{12}C$ and $^{3}He$ induced reactions. In section II, we discuss statistical model calculations and compare between the calculated spectra from 'CASCADE' and 'GEMINI' codes. The analysis of the experimental results has been presented in section III. Discussions of the results and a conjecture have been presented in section IV. Finally the conclusion has been given in section V.

## II. STATISTICAL MODEL PREDICTIONS

One of the standard methods of the measurement of the temperature of the ensemble of residual nuclei (at the instant of the break-up of the exit channel dinuclear system comprising residual and the emitted fragment) is by measuring the slope of the exponential tail of the spectrum of the emitted particles, provided the decay is statistical. The statistical evaporation spectrum of fragments emitted from a compound nucleus can be written [5,6] as

$$P(x) \propto \exp\left(-\frac{x}{T}\right) erfc\left(\frac{p-2x}{2\sqrt{pT}}\right) \tag{1}$$

$$x = E_{kin}(c.m.) - V_C.$$

Here $E_{kin}(c.m.)$, p, T, P(x) are the exit channel center of mass kinetic energy, amplification parameter, temperature of the ensemble of the residual nuclei and the corresponding probability of the emission of the particle, respectively. $V_C$ is a parameter that is equal to the Coulomb barrier for the zero orbital angular momentum of the system [5]. Usually for low-energy nuclear reactions, the temperature is obtained by fitting evaporation proton and alpha spectra from an ensemble of compound nuclei with eq. (1) [6,7]. It was found [6,7] that high-statistics alpha spectra can be fitted within a few percent using an one source term as given in eq. (1) and that an average temperature of the ensemble of the residual nuclei can be extracted. The effect of the sequential decay can be considered by adding up several source terms (like eq. (1)) with decreasing temperatures [8]. The inclusion of additional two or three source terms with decreasing temperatures can fit experimental spectra within 0.5% [8]. It should be possible to use the heavier fragment (such as Li, Be, B, C) evaporation spectra to determine the temperatures of the corresponding ensembles of the residual nuclei, provided those emissions are statistical. At relatively lower excitation energy when the orbital kinetic energies taken away by the heavier fragments remain significant compared to the excitation energy of the compound system, the temperatures of the residual nuclei obtained by fitting heavier fragment evaporation spectra are generally expected to be lower compared to those obtained from proton or alpha spectra.

We used the statistical model code 'CASCADE' [2] to calculate the spectra of the neutron, proton, alpha and heavier fragments from $^{16}O+^{89}Y$ reaction at $E_{Lab}(^{16}O)$ =96 MeV. In the 'CASCADE' code, the neutron, proton and alpha particles were the main particle emission channels and the lithium or boron or carbon particle was used as a fourth particle channel one at a time. The 'CASCADE' code executed typically 35 sequential decay steps until the cross-section of the evaporation residues fell below 1 mb. The final summed evaporation spectra of the alpha, lithium, beryllium, boron and carbon particles obtained from the calculations were fitted with eq(1) to obtain the corresponding slope temperatures which should indicate the average temperatures of the corresponding ensembles of the residual nuclei. The 'GEMINI' code [1] calculations were done to compare with the 'CASCADE code calculations. In the 'GEMINI' code, neutron, proton, alpha and lithium particle emissions were calculated using Hauser-Feshback transition state formalism. The emission of heavier fragments such as beryllium, boron and carbon were treated as evaporation emissions simultaneously. The emission of the fragments from the sequential decay of the pre-fragments (those are produced in the initial binary decay of the compound nucleus) is included in the 'GEMINI' code. The 'GEMINI' code gave relative cross-sections for the emission of different ions and the spectra were normalized with respect to the corresponding calculated spectra from the 'CASCADE' code to compare the corresponding spectral shapes calculated from the two codes. The calculated spectral shapes of alpha, lithium, beryllium, boron and carbon obtained from the 'CASCADE' agree very well with the corresponding spectral shapes calculated by the 'GEMINI' code. In Fig. 1, we show the calculated alpha spectra (in the center of mass frame) for both the 'CASCADE' and 'GEMINI' codes and a fit using eq. (1). The fit yields T=2.9 MeV, p=4.0 MeV and $V_C$=12.5 MeV. In Fig. 2-5, we show the calculated $^6$Li, $^9$Be, $^{11}$B and $^{12}$C spectra using the 'CASCADE' and 'GEMINI' codes and fits to corresponding 'CASCADE' calculations using eq. (1). The results of the fits have been presented in Table 1. It was found that the temperatures of the residual nuclei generally decrease for the emission of the heavier fragments as expected at such low excitation energy ($E_X$=76 MeV). On the other hand, the value of the amplification parameter (p) increases from 4.0 MeV for the alpha particle to 20.0 MeV for the carbon particle.

The calculations have been done using the level-density parameter a=A/8 and the extracted parameters (T and p) are essentially independent of the transmission coefficients, deformation parameter, critical angular momentum etc. used in the calculations. The use of different optical-model parameters resulting in a different set of transmission coefficients and different critical angular momenta change the calculated cross-sections of the heavy fragments considerably, but hardly affect the slope of the exponential tail of the spectrum and the corresponding extracted temperature, because the distribution of the excitation energy of the residual nuclei essentially remains independent of the transmission coefficients of the ejectiles. So we have normalized the calculated statistical-model spectra of heavy ions (Li, Be, B, C) with respect to the experimental spectra and compared the corresponding spectral shapes. The statistical model calculations have also been done considering the emission of heavy ions in their first and second excited states and the corresponding spectral shapes remain essentially unchanged. The qualitative features of the calculations showing substantial lower temperature from the heavy fragment spectra is model independent and essentially follows from the fact that the heavy fragments take away a significant fraction of the available energy as their kinetic energies and also because the masses of the residual nuclei are not significantly lower than the parent compound nucleus.

The 'CASCADE' code [2] calculations were done for $^3$He+Ag reaction at E($^3$He)=90 MeV and $^4$He, $^6$Li, $^9$Be, $^{11}$B and $^{12}$C evaporation spectra were generated. The 'GEMINI' code [1] calculations were also done for the reaction and the evaporation spectra were calculated. The spectral shapes of the evaporation particles obtained from the 'GEMINI' code agree very well with the corresponding spectra obtained from the 'CASCADE' code calculations. In Fig. 6, we show the calculated alpha spectra (in the center of mass frame) for both the 'CASCADE' and 'GEMINI' codes and a fit to 'CASCADE' calculation using eq. (1). The fit yields T=3.0 MeV, p=2.0 MeV and $V_C$=13.5 MeV. In Fig. 7-9, we show the calculated $^6$Li, $^{11}$B and $^{12}$C spectra using the 'CASCADE' and 'GEMINI' codes and fits using eq. (1). As before, the spectra obtained from the 'GEMINI' code have been normalized to overlay with the spectra obtained from the 'CASCADE' code. The results of the fits have been presented in Table 1. It was found that the temperatures of the residual nuclei drop continuously for the emission of the heavier fragments as expected at this low excitation energy ($E_X$=102 MeV). On the other hand, the value of the amplification parameter (p) increases from 2.0 MeV for the alpha particle to 8.5 MeV for the carbon particle. The calculations have been done using the level-density parameter a=A/8 and the extracted parameters are essentially independent of the transmission coefficients used in the calculations. The statistical model ('CASCADE' and 'GEMINI') calculations were also performed for $^3$He+Ag reaction at E($^3$He)=198.6 MeV. The temperatures obtained from the calculated $^4$He,

$^6$Li, $^9$Be, $^{11}$B and $^{12}$C spectra have been found to be about the same in all cases ($\approx$ 4MeV). This is because the kinetic energies carried away by the heavy fragments are small compared to the excitation energy of the compound nucleus ($E_X$= 212 MeV) and so the extracted temperatures from alpha and heavy ion fragments are about the same. The statistical model calculations were also performed for $^{16}$O+$^{93}$Nb reaction at $E_{Lab}$($^{16}$O)=116 MeV and the evaporation spectra of alpha, lithium, beryllium, boron and carbon were generated using the statistical model codes — 'CASCADE' and 'GEMINI'. The calculated spectra were fitted with eq.(1) and the extracted temperature (T) and p-parameters have been presented in Table 1. According to the 'GEMINI' code, most of the lithium fragments come from the sequential decay of the excited pre-fragments such as beryllium and boron, whereas 'CASCADE' code only calculates evaporation lithium spectrum. We find that the slope parameters (T) extracted from the lithium spectra calculated by the code 'GEMINI' are similar to those extracted from the corresponding calculated evaporation lithium spectra using the 'CASCADE' code.

## III. ANALYSIS OF EXPERIMENTAL DATA

In order to test these predictions, we used both the literature data [3, 4] and data from our own experiments. We performed experiments forming [9] the same compound nucleus $^{105}$Ag at the same excitation energy ($E_X$=76 MeV) and with very similar spin distributions ($\ell_{crit}$ equal within 10%) by $^{16}$O+$^{89}$Y reaction at E($^{16}$O)$_{Lab}$=96 MeV and $^{12}$C+$^{93}$Nb reaction at E($^{12}$C)$_{Lab}$=85.5 MeV. The details of the experiment are given in ref [9]. In the center of mass frame, the back-angle angular distributions of alpha, lithium, beryllium, boron, carbon etc. showed back-angle rises. The angular distributions of the heavier fragments such as carbon can be approximated by a 1/sin$\theta_{c.m.}$ function [9]. The angular distribution of lighter particles such as lithium can be fitted with a (a+b cos $^2\theta$) function as shown in Fig. 10. This kind of back-angle rise of the angular distributions for the lighter ions and approximate 1/sin$\theta_{c.m.}$ angular distribution for the heavier ions such as carbon are characteristics of statistical emission of the fragments [3,5] from an equilibrated compound nucleus. We have also studied the angle-integrated ratios of the yields of boron to carbon, beryllium to carbon and lithium to beryllium as a function of the exit channel excitation energy for both $^{16}$O+$^{89}$Y and $^{12}$C+$^{93}$Nb reactions and find [Fig. 11, Fig. 12] that the ratios of the yields of boron to carbon, beryllium to carbon and lithium to beryllium for $^{16}$O+$^{89}$Y and $^{12}$C+$^{93}$Nb reactions forming the same composite at the same excitation energy and similar spin distribution overlap with each other reasonably well, implying no significant entrance channel effect and hence their statistical origin from an equilibrated compound nucleus. The corresponding ratios did not overlap exactly due to the mismatch of the spin distributions of the compound nuclei formed by $^{16}$O+$^{89}$Y and $^{12}$C+$^{93}$Nb reactions. In Fig. 13, we show the overlayed lithium spectra at different angles from $^{12}$C+$^{93}$Nb and $^{16}$O+$^{89}$Y reactions after suitable normalizations. We find essentially identical spectral shape at different angles implying statistical emission of lithium particles at back-angles. Similar results have also been obtained for other ions. The angular distributions show back-angle rise for all ions and tend to approximate 1/sin$\theta_{c.m.}$ function for heavier fragments such as carbon. So we have established that the emissions of the heavier fragments such as lithium, beryllium, boron and carbon from $^{16}$O+$^{89}$Y and $^{12}$C+$^{93}$Nb reactions at back-angles (center of mass frame) are statistical.

In Fig. 14, we have overlayed the calculated (using the 'CASCADE' code) and the experimental alpha spectra from $^{16}$O+$^{89}$Y reaction at E($^{16}$O)$_{lab}$=96 MeV and find that the shapes of the calculated and experimental alpha spectra are almost identical. A one source fitting of the experimental alpha spectrum (using eq.(1)) has been shown and the extracted temperature agrees very well with that obtained from the calculated statistical spectrum (Table 1). The fitting of the alpha spectra with eq. (1) was discussed earlier [8] and it was found that a one source fit was reasonable, although a 3-source fit with decreasing source temperature was definitely better. In Fig. 15-18, the experimental angle-integrated lithium, beryllium, boron, carbon and the corresponding calculated statistical model spectra have been overlayed. The absolute normalisations of the spectra have been obtained from the measured Rutherford elastic cross-section at a suitable forward angle. The theoretical spectra have been normalised and shifted with respect to the corresponding experimental spectrum to overlay their peak positions. The experimental spectra have been fitted with eq. (1) and T and p-parameter have been extracted for each case. We find that the low energy part of the heavy ion spectra matches well, but the slopes of the calculated spectra are significantly steeper than the experimental spectra. The largest slope anomaly is seen for the lithium spectrum. In Table 1, we show the fitted values of T and p obtained from statistical model calculations and the experimental spectra. We get very similar results from the study of alpha, lithium, beryllium, boron and carbon particles emitted at back-angles from $^{12}C+^{93}Nb$ reaction at $E_{Lab}(^{12}C)$=85.5 MeV forming the same compound nucleus $^{105}Ag$ at excitation energy = 76 MeV with very similar spin distribution. The first question is whether the presence of possible impurities in the target might cause the observed distortion of the slope of the back-angle spectra. The possible low Z contaminants in the target are carbon and silicon that might come from the pump oil and oxygen. We did not see any observable presence of low Z contaminants in the elastic yield measured by monitor detectors placed at forward angles. Moreover the kinetic energy of the emitted lithium and heavier fragments emitted at back-angles from the reaction of oxygen with low Z contaminants (carbon, oxygen, silicon) will be very low and they cannot possibly produce any significant effect in the high energy tail of the fragments emitted at back-angles from $^{16}O+^{89}Y$ and $^{12}C+^{93}Nb$ reactions. They might slightly contaminate the low energy part of the observed spectra, but no significant anomalies have been observed in the lower energy side of the heavy fragment spectra. So the effect of any such low Z contaminant can be neglected. High Z contaminants like tantalum and dysprosium might be present in the target at a level of < 100 ppm (as per foil supplier's catalogue). However the cross-sections of lithium and heavier fragments emitted at back-angles from the reaction of oxygen with such high Z-contaminants should be orders of magnitude lower than the corresponding cross-sections from $^{16}O+^{89}Y$ reaction, because of the Coulomb barrier effect. Moreover such high Z-contaminants should produce colder spectra. So the observed anomaly of the high energy tail of the emitted heavy fragments from $^{16}O+^{89}Y$ and $^{12}C+^{93}Nb$ reactions cannot be because of the presence of the small amount of contaminants in the target.

It might be argued that the observed mismatch of the slopes of the calculated and experimental spectra indicates non-statistical emission of the fragments in the high energy tail region. However the angular distributions of the tail region of the spectra are not different from the angular distribution of the total spectra and Fig. 13 shows that the spectral shapes of lithium particles are almost identical at different angles. We do not see any significant entrance channel dependence in the high energy tail region of the heavy ion spectra from Fig. 11 and Fig. 12. So there is no indication of any significant non-statistical emission process in the high energy tail region.

There might be a question whether orbiting reaction (10, 11) might be responsible for the observed slope anomaly, because the angular distributions of the emitted fragments show back-angle rise in the case of orbiting reactions also. However in the case of orbiting processes, strong entrance channel effect was seen [9, 10] and that is absent in this case [Fig. 11 and Fig. 12]. Moreover in the case of orbiting reaction, although all the degrees of freedom (such as the shape of the orbiting composite) are not equilibrated, temperature equilibration is attained. So the spectral shape of the orbiting fragments is expected to be similar [11] to that from the statistical model calculations, although the absolute yield of the orbiting fragments is much higher than the estimates from the statistical model calculations. There is no known reaction mechanism that can produce back angle rise of the angular distributions of the emitted fragments, but the nuclei fail to attain at least thermal equilibration. The thermal equilibration implies that the spectral shape of the emitted fragments would be similar to that expected from the statistical model calculations. So the observed slope anomaly (Fig. 15-18) cannot be explained by orbiting or similar reaction mechanisms.

In order to further check whether the observed slope anomaly of the back-angle heavy ion spectra from $^{16}O+^{89}Y$ and $^{12}C+^{93}Nb$ reactions are characteristics of those particular reactions or a more general feature of the statistical heavy ion emission process, we have compared back-angle heavy ion evaporation spectra from $^3He+Ag$ reaction at $E_{Lab}(^3He)=90$ MeV [3] with the 'CASCADE' and 'GEMINI' code calculations. The temperature (T=3.0 MeV) and p-value (p=2.0 MeV) obtained from the calculated $^4He$ spectrum [Fig. 6] agree very well with the corresponding parameters obtained from the experimental $^4He$ spectrum of $^3He+Ag$ reaction at $E(^3He)=90$ MeV taken at back-angle [7]. The angular distribution [3] of the emitted heavy ions (lithium, beryllium, boron and carbon) from $^3$He+Ag reaction at E($^3$He)$_{lab}$=90 MeV showed back-angle rises and they tended to an approximate 1/sinθ$_{c.m.}$ function for the heavier fragments. As discussed in ref [3], these kinds of angular distributions at back-angles indicate the statistical origin of those fragments. In Fig. 19-21, we show overlayed plots of calculated lithium, boron and carbon evaporation spectra along with the corresponding experimental spectra taken at back-angle from ref [3]. The experimental spectra have been fitted with eq. (1) and the values of T and p-parameter have been extracted for each case. As before, we find that the slopes of the calculated heavy fragments are significantly steeper than the corresponding slopes obtained from the experimental spectra and the slope anomaly is the largest for the lithium spectrum. The results have been tabulated in Table 1. Pre-equilibrium emission of neutrons is expected from this reaction and it produces a broad distribution of the excitation energy of the compound nucleus around a mean value of ~82 MeV [3]. The statistical model calculations do not consider any pre-equilibrium neutron emission and so the temperature extracted from the calculated statistical model spectra should give a upper limit of the temperature. The consideration of a distribution of excitation energy around 82 MeV should increase the slope anomaly. At higher excitation energy (150-200 MeV), reasonable agreements [4] with the statistical model calculations have been obtained. The shapes of the heavy ion spectra (beryllium, boron, carbon) emitted at back-angles from the $^3$He+Ag reaction at E($^3$He)=198.6 MeV agree reasonably well [4] with the corresponding spectra calculated from the statistical model codes. The temperatures were obtained by fitting the spectra with eq. (1) and temperatures around T ≈ 4MeV were obtained from both the alpha and heavy ion spectra. In Fig. 22, we show the overlayed calculated ('GEMINI') and experimental boron spectra for $^3$He+Ag reaction at E($^3$He)=198.6 MeV [4]. The calculated and experimental spectra match reasonably well, showing that the slope anomaly disappears at high excitation energy.

We also performed an experimental study of the back-angle heavy ion emission from $^{16}$O+$^{93}$Nb reaction at E($^{16}$O)$_{Lab}$=116 MeV. A 5 pnA $^{16}$O beam from the Variable Energy Cyclotron Center, Kolkata at E($^{16}$O)$_{lab}$ = 116 MeV was used to bombard a 1mg/cm$^2$ thick $^{93}$Nb foil and alpha, lithium, beryllium, boron and carbon spectra were recorded. The details of the experiment were given in ref [9]. In Fig. 23 and 24, we show overlayed plots of calculated and experimental alpha and lithium spectra. The spectral shapes of the alpha spectra match reasonably well. However as before the slope of the calculated lithium spectrum is significantly steeper compared to that of the experimental spectrum. In Fig 25-27, we show the overlayed plots of the calculated and experimental (angle-integrated) beryllium, boron and carbon spectra. We find from Fig. 24-27 that the slope of the calculated spectrum is always steeper than the corresponding experimental spectrum and the deviation is in the tail region, although there is reasonable agreement between the calculated and experimental spectra in the rest of the spectrum. The extracted slope parameter (T) of the calculated spectrum has always been found to be lower than the corresponding T-parameter extracted from the experimental spectrum as shown in Table 1. In the case of the overlayed experimental and statistical model spectra (Fig. 14-27), the calculated spectra were typically shifted by 0.5 – 1.5 MeV to match with the peak positions of the corresponding experimental spectra. Larger shifts (~4 MeV) of the calculated beryllium and boron spectra towards the lower energy were required for $^{16}$O+$^{89}$Y and $^{16}$O+$^{93}$Nb reactions implying deformation of the residual nuclei. No normalization factor was required for calculated alpha and carbon spectra, however large normalization factors (factor of ten or more) were required for calculated beryllium and boron spectra, because the optical model potentials required to calculate the transmission coefficients of beryllium and boron are poorly known. The absolute cross-section of the emitted fragments depends very strongly on the transmission coefficients and it is possible to change the absolute cross-sections of the weak channels by orders of magnitude by choosing different optical model potentials for calculating the transmission coefficients. However the spectral shapes (particularly the T-parameter) remain about the same for different sets of the transmission coefficients. So we have emphasized the comparison of the corresponding spectral shapes. The experimental spectra were fitted with eq. (1) and T and p-parameters were extracted for each case. As before, we find that the slopes of the calculated spectra are steeper than those obtained from the experimental spectra, but the slope anomaly has decreased somewhat compared to lower energy data taken at $E_{Lab}(^{16}O) = 96$ MeV. In Table 1, we have presented the fitted values of T and p for both the statistical model and the experimental spectra. In Fig. 23-27, both the 'CASCADE' and 'GEMINI' calculations producing similar spectral shapes have been shown.

We find from our study of the back-angle alpha and heavy ion spectra from $^{16}O+^{89}Y$ reaction at $E_{Lab}(^{16}O) = 96$ MeV, $^{12}C+^{93}Nb$ reaction at $E_{Lab}(^{12}C)=85.5$ MeV, $^{16}O+^{93}Nb$ reaction at $E_{Lab}(^{16}O)=116$ MeV and $^{3}He+Ag$ reaction at $E_{Lab}(^{3}He)=90$ MeV and 198.6 MeV that the experimental and calculated alpha spectra (from the statistical models) agree well in all the cases. However the experimental heavy ion spectra show significantly gentler slope than the calculated statistical spectra. This anomaly is largest for the lithium spectra and decreases as the excitation energy of the compound nucleus increases. The qualitative features of this anomaly are the same for $^{16}O$, $^{12}C$ and $^{3}He$ induced reactions. The observed back-angle rise of the angular distribution and the lack of any entrance channel dependence of the reaction products imply the statistical origin of these reaction products. If reaction dynamics is responsible for the slope anomaly, the anomalies should be qualitatively different for $^{16}$O and $^{3}$He induced reactions. The observation of a very similar slope anomaly for both the $^{16}$O and $^{3}$He induced reactions indicate that the reaction dynamics should not be responsible for the anomalies.

**IV. DISCUSSION**

The shapes of the calculated statistical model spectra are very robust and do not depend on deformation parameters, transmission coefficients, critical angular momentum etc. The level-density parameter of the statistical model is essentially the only parameter that can change the slope of the fragment spectra. It is known [1] that there is no excitation-energy dependence of the level-density parameter in the A ≈100 mass region. Even if we introduce any excitation-energy dependence of the level-density parameter, it cannot explain the observed features — no anomaly for the alpha spectrum, very significant slope anomaly for the lithium spectrum and somewhat smaller anomalies for other heavier particle spectra. An additional angular momentum dependence of the nuclear level density can be introduced by multiplying the liquid drop moment of inertia by a factor. If we multiply the liquid drop moment of inertia by 0.9, then the cross-section of $^{12}$C increases by an order of magnitude, but the slope of the $^{12}$C spectrum becomes only a little bit gentler increasing the extracted temperature from 1.7 to 2.2 MeV. The corresponding effects on the cross-sections and the extracted temperatures from the boron, beryllium, lithium and alpha spectra become progressively smaller as those particles carry progressively lower orbital angular momentum. So such angular momentum dependence of the nuclear level density cannot explain why the slope anomaly is largest for the lithium that carries much lower orbital angular momentum than $^{12}$C, whereas there is no anomaly for the alpha spectrum. So we think that the observed slope anomalies found from both the literature data [3] and our data cannot be explained by adjusting the parameters of the statistical model or by reaction dynamics.

There is also a question whether the slope anomaly might be due to the production of the observed fragments from the sequential decay of the pre-fragments. According to the 'GEMINI' code, such effect should be largest for the lithium spectra. The 'GEMINI' code calculations have been done considering such effect and no significant change of the slope parameter (T) has been found compared to the corresponding (CASCADE code) evaporation spectrum (Fig. 2, Fig. 7, Fig. 19, Fig. 24). Moreover such sequential decay of the pre-fragments should increase at higher excitation energy. So if the observed slope anomaly is due to the sequential decay of the pre-fragments, then the disagreement between the experimental slope parameter (T) and that obtained from the calculated evaporation spectrum ('CASCADE' code) should increase at higher excitation energy contrary to the observations.

**Effect of Shape Polarization**

Let us examine if Moretto's shape polarization model [5] for the emission of large fragments from a compound nucleus might explain the observed anomalies. On the basis of this model, Moretto deduced [5] deduced eq. (1) describing the shape of kinetic-energy spectrum of the emitted heavy-ion spectra from a compound nucleus. The equation contains three parameters p, T and $V_C$ and they are inter-related by the relation $\Delta V_C = 2\sqrt{pT}$, where $\Delta V_C$ denotes the fluctuations of the Coulomb barrier. However the model does not give any ab-initio method to calculate p and T parameters. We fitted our experimental and statistical model spectra with eq. (1) and extracted T, p and $V_C$ parameters. There is interplay between T and p parameters and the decrease of p parameter can increase T parameter. We performed best fits by minimizing corresponding chi-square values and checked that if we decrease p by a significant amount and try to fit the spectra by increasing T, then no reasonable fit can be obtained for lithium, beryllium, boron and carbon spectra.

In order to study the effect of shape polarization, let us first assume that both the residual and ejectile nuclei are spherical and the distance between the centers of the two nuclei is r when they are in a touching configuration. So we obtain $V_C = \frac{Z_{res}Z_{ejec}}{r}$, where $Z_{res}$ and $Z_{ejec}$ denote the atomic numbers of the residual and ejectile nuclei respectively. Then (assuming no change of the spherical shape), we obtain $\Delta V_C = -\frac{Z_{res}Z_{ejec}}{r^2}\Delta r$. Following ref [5, 7], we assume $\Delta r \propto \sqrt{T}$ (considering r as the deformation parameter) and then obtain using the equation $\Delta V_C = 2\sqrt{pT}$ the following relationship

$$p \propto \frac{(Z_{res}Z_{ejec})^2}{r^4}. \quad\quad\quad\quad\quad\quad\quad\quad\quad\quad\quad\quad\quad\quad\quad\quad (2)$$

From eq. (2), we obtain that the ratio of the p-parameters for carbon ($p_C$) and helium ($p_{He}$) spectra should be 5.25 for $^{16}O+^{89}Y$ reaction, whereas the corresponding ratios extracted by fitting the experimental and statistical model spectra are = 3.75±0.5 and 4.5 respectively. In the case of emission of the lithium particles from $^{16}O+^{89}Y$ reaction, we obtain from eq. (2), ($p_{Li}/p_{He}$)= 1.9, whereas the corresponding ratios obtained from both the experimental and statistical model spectra are 1.5± 0.2 and 1.5 respectively. In the case of $^3He+Ag$ reaction, the calculated ratio of $p_C/p_{He}$=4.88 using eq. (2), whereas the corresponding ratio obtained by fitting the experimental and statistical model spectra are = 5.45±0.75 and 4.25 respectively. We think that the increase of p-value (for heavier ejectiles) as obtained by fitting the experimental spectra with eq. (1) is in qualitative agreement with our simple estimates from eq. (2). Somewhat lower values of the ratios obtained from the experimental spectra of $^{16}O+^{89}Y$ reaction compared to the estimates from eq. (2) might be due to the shape polarization effect that should reduce the value of p parameter compared to the estimates of eq. (2) that is based on the assumption of spherical nuclei with no shape polarization. So shape polarization should reduce the value of p-parameter (from the assumption of spherical shape) for large fragments and there might be some qualitative evidence for such an effect from the experimental spectra of heavier fragments from $^{16}O+^{89}Y$ reaction. Let us now examine possible effect of the shape polarization on the spectral temperature (T). According to the shape polarization model of Moretto [5], the temperature (T) is proportional to the square of the fluctuation of the deformation co-ordinate i.e. $T \propto (\Delta r)^2$. The fluctuation of the deformation co-ordinate will certainly be larger for the emission of the larger fragments. So the shape polarization effect should tend to increase the spectral temperature obtained from the spectra of the larger fragments. Hence according to this argument, shape polarization effect should be more important for the carbon emission than lithium emission. However, experimentally, we have found that the slope anomaly is largest for the lithium spectrum and highest spectral temperature has been extracted from the lithium spectrum. The effect of shape polarization should not be less important at higher excitation energy. However it has been observed that the slope anomaly decreases rapidly at higher excitation energy and it is not visible around $E_X \sim (150\text{-}200)$ MeV. So we think that the inclusion of the shape polarization effect in a statistical model cannot explain the observed results.

**Classical fission delay due to nuclear viscosity**

Another point to examine is whether the time delay of the emission of large fragments from the compound nucleus (similar to fission delay) because of the effect of the viscosity of the nuclear medium might affect the spectral temperature of the fragment. The observation of relatively long fission time ( ~$10^{-20}$ sec) of the excited high Z compound nuclei from neutron multiplicity measurements led to the speculation that the viscosity of the hot nuclear medium might be responsible [12-14] for slowing down evolution from the equilibrated compound nuclear shape to the scission point. At higher excitation energy, the hypothesis requires much higher values of the viscosity parameter to explain neutron pre-scission multiplicity data. Recently McCalla and Lestone [15] proposed a different model without increasing the viscosity parameter steeply with the temperature for relatively lower Z nuclei (Po) where the fission barrier is not very low. However direct measurements by X-ray and crystal blocking techniques have shown very long fission delay times (~$10^{-18}$ sec) even for the highly excited uranium-like and trans-uranium nuclei [16-18). These results have not so far been explained by the viscosity effect.

If we consider that the emissions of the fragments like lithium, beryllium, carbon etc. from $^{16}O+^{89}Y$ and $^3He+Ag$ reactions are delayed by the viscosity effect slowing down the evolution of the equilibrated compound nucleus to scission point, then the spectral temperatures of the fragments should be lower than the expectations from the standard statistical model predictions, because of the emission of pre-scission neutrons during the long evolution time period. So the consideration of any such classical time delay effect should certainly produce colder spectrum compared to the standard statistical model predictions, contrary to our observations. So the observed higher spectral temperatures of the fragments compared to the statistical model predictions cannot be explained by classical time delay effects (such as due to the nuclear viscosity), because such effects should produce lower temperature, because of the emission of the pre-scission neutrons.

**Quantum Mechanical Analysis (Compound Nuclear Wave Function)**

In order to understand the observed slope anomaly, let us examine the statistical model from a quantum mechanical perspective. The wave function of a statistical compound nucleus is a linear superposition of the wave functions of all possible exit channel dinuclear states, where each dinuclear state wave function (a quasi-bound state of a heavy-ion fragment and residual nucleus) is undergoing exponential decay in time with a Breit-Wigner width. Let $\psi_{CN}$ denotes compound nucleus wave function and $\psi_k$ that of a dinuclear state comprising $k^{th}$ residual nucleus (with excitation energy $\varepsilon_{x1}$) and the corresponding fragment nucleus (with excitation energy $\varepsilon_{x2}$), then

$\psi_{CN} = \sum_k \sum_{\varepsilon_{x1}, \varepsilon_{x2}} \sum_{L,S} a_k(\varepsilon_{x1}, \varepsilon_{x2}, L, S) \psi_k(\vec{r}, t, L, S, \varepsilon_{x1}, \varepsilon_{x2})$, where $|a_k(\varepsilon_{x1}, \varepsilon_{x2}, L, S)|^2$ denotes phase space factor for the corresponding dinuclear state. L and S denote the corresponding orbital angular momentum and channel spin respectively. Considering exponential decay of the dinuclear state, the wave function $\psi_k$ can be written as

$\psi_k(\vec{r}, t, L, S, \varepsilon_{x1}, \varepsilon_{x2}) = \psi_k(\vec{r}, L, S) \exp\left(\frac{-\Gamma_k t}{2\hbar}\right) \exp(iE_k t/\hbar)$, where $E_k$ denotes the energy of the $k^{th}$ dinuclear state, t is time and $\Gamma_k$ denotes the Breit-Wigner width of the $k^{th}$ dinuclear state. So the time dependence of the compound nucleus wave function can be obtained from the above equations. In any statistical model calculation, concerned Breit-Wigner widths are assumed to be zero and then the entropy function and the corresponding temperature of the exit-channel dinuclear system become approximately proportional to the square root of the thermal energy of the exit-channel dinuclear system. However the Breit-Wigner width of an exit-channel dinuclear state might not be negligible for heavy-ion emissions. It should depend on the stability of the heavy-ion fragment in the vicinity of the residual nucleus and also be proportional to the transmission coefficient of the fragment. An alpha particle in the presence of a residual nucleus is not expected to break apart and so the Breit-Wigner width of a dinuclear state comprising an alpha particle and the residual nucleus should only be proportional to the transmission coefficient of the alpha particle and might be approximated as a sharp state, as assumed in the statistical models. However a dinuclear state comprising a $^6$Li and the residual nucleus should be very short-lived, because $^6$Li should promptly break apart in the nuclear and Coulomb field of the residual nucleus. Similarly the dinuclear states comprising beryllium and the residual nucleus, boron and the residual nucleus and carbon and the residual nucleus should also be short-lived compared to that comprising an alpha and the residual nucleus. Moreover the angular momentum averaged transmission coefficient of the heavier fragments should be larger than that for the alpha particles, thus increasing the Breit-Wigner width for the dinuclear states comprising heavy fragment and the residual nucleus.

**Conjecture**

In the statistical model calculation, the entropy ($S \propto 2\sqrt{aE}$, $a$ being the level density parameter) of the nuclear system at energy E is obtained assuming a sharp (non-decaying) energy state. We conjecture that the effect of the Breit Wigner widths of the states might be considered by convoluting the standard entropy function with a Breit-Wigner function and then calculating the temperature of the system by the standard method. So for a decaying nuclear system having Breit-Wigner width Γ around a mean energy U, entropy and the temperature might be written as

$$\langle S \rangle \approx \frac{\int_{-\infty}^{\infty} \frac{2\sqrt{aE}}{(E-U)^2 + 0.25\Gamma^2} dE}{\int_{-\infty}^{\infty} \frac{1}{(E-U)^2 + 0.25\Gamma^2} dE} \quad \text{and} \quad \frac{1}{T} = \frac{\partial |\langle S \rangle|}{\partial U} \quad \quad \text{.............(3)},$$

Using eq. (3), a plot of the temperature (T) versus the thermal energy (U) for various values of Breit-Wigner width ($\Gamma$) is shown in Fig. 28 for A=93 and level density parameter = A/8. The standard statistical model result corresponds to $\Gamma$=0. We find that for large values of $\Gamma$, the calculated temperature becomes very large for small values of U, but as U increases, the calculated temperatures comes down quickly and approaches the statistical model values. In the case of the emission of the lithium fragments from the compound nucleus $^{105}$Ag ($E_X$=76 MeV), we are dealing with the average thermal energy of the residual nuclei (U~35-40 MeV) and in order to explain our slope anomaly corresponding to the experimentally obtained values of T, we shall need $\Gamma$ ~100 MeV. Somewhat smaller values ($\Gamma$ ~50-80 MeV) would be required to explain slope anomalies of other heavy ion spectra. These results imply prompt breakup of lithium, beryllium etc. in the vicinity of the nuclear and Coulomb field of the residual nucleus. Since the lithium nucleus is expected to break apart most promptly compared to carbon, boron or beryllium, the Breit-Wigner width of the dinuclear state comprising lithium and the residual nucleus should be the largest resulting in the highest temperature of the lithium spectrum and largest departure from the standard statistical model prediction (based on sharp Breit-Wigner states), as observed experimentally. On the other hand, the temperature of the alpha spectrum should agree with the prediction of the standard statistical model, because of the expected small Breit-Wigner width of the dinuclear state comprising an alpha and the residual nucleus due to the stability of alpha nucleus in the vicinity of the nuclear and Coulomb field of the residual nucleus. Numerical estimates also show (Fig. 28) that at higher (thermal) excitation energy, the effect of the Breit-Wigner width on the temperature drops very rapidly and the calculated temperature tends to approach the statistical model prediction. So this conjecture regarding convoluting the entropy function with a Breit-Wigner function to calculate the temperature of the residual nucleus has the potential to explain the observed anomaly.

Of course, there is an associated problem with this conjecture. The question is how lithium and other particles are coming out from the close proximity of the residual nuclei and reaching detectors satisfying the condition of long lifetime as required by the statistical-emission process [19-21], if indeed they have such a large Breit-Wigner width corresponding to prompt break-up in the nuclear and Coulomb field. A possible resolution of the puzzle could be the initial quantum mechanical delay (non-exponential decay time scale ~$10^{-18}$ sec – $10^{-21}$ sec for nuclear systems) [22-25] of the dinuclear states resulting in a long survival time and statistical characteristics.

**Quantum Mechanical Delay**

The decay of an unstable quantum state should initially be non-exponential due to the quantum mechanical probability of regenerating the initial state from the decay products [22-25]. According to the exponential decay law, the survival probability [P(t)] of a decaying state after time t (t ~0) is given by
$P(t) \propto (1 - \Gamma t)$, whereas quantum mechanics predicts that the corresponding survival probability [23-25] should be $P(t) \propto [1 - (const \times t)^2]$. So quantum mechanics predicts an approximately flat initial survival probability (close to unity) for sometime, because quantum mechanical unitary time evolution is time-reversible and cannot by itself lead to time-irreversible exponential decay process. However the quantum mechanical reversible decay process certainly cannot persist for a long time, because long-time exponential behavior of a decaying system is experimentally very well established. So we may think (approximately) that the exponential decay of an unstable state starts after a quantum mechanical time delay. This non-exponential decay time scale or quantum mechanical delay time depends on the system concerned and its estimate is model dependent. Recently Wilkinson et al. [26] observed non-exponential decay (flat survival probability at early time) in quantum tunnelling of an atomic system. So far nonexponential decay of the radioactive nuclei has not been seen [27] because of their expected very short time scale of the order of $\geq \hbar/$(energy release) [23-25]. However the nonexponential nuclear decay time scales might be comparable to the lifetime of highly excited compound nuclei and play a role in their decays. As a result of this quantum mechanical delay (nonexponential decay time scale), an unstable quantum state with a large Breit-Wigner width (which should result in a very short lifetime) is expected to live for a significantly longer time. Since lithium and other emitted fragment nuclei are bound states of nucleons, so their exponential decay (in the presence of the residual nucleus) should also start after a quantum mechanical delay, thus allowing them to form a long-lived dinuclear state with the residual nucleus and come out before decaying, provided the quantum delay (non-exponential time scale) associated with the fragment breakup is longer than that associated with the corresponding dinuclear state of fragment and the residual nucleus, as expected.

## V. CONCLUSION

We have studied experimentally the statistical emission of alpha and heavier fragments produced in low energy (4-8 MeV/A) $^{16}$O and $^{12}$C induced nuclear reactions at back-angles. We have compared our experimental fragment spectra and the existing back-angle experimental statistical fragment spectra from $^3$He+Ag reaction at $E_{Lab}(^3He)$=90 MeV and $E_{Lab}(^3He)$=198.6 MeV [3,4] with the statistical model calculations. The observed shapes of the alpha particle spectra agree with the corresponding statistical model calculation in all the cases. However the temperatures extracted from the slopes of the spectra of the heavier fragments are significantly higher than the expectations from the statistical model. The anomaly is largest for the lithium spectrum and all the slope anomalies decrease at higher excitation energy for all the reactions studied. The observed similarity of the slope anomalies for two very different entrance channels ($^{16}$O and $^3$He induced reactions) indicates that the reaction dynamics is unlikely to be the reason behind the observed slope anomalies. The anomalies cannot be understood by adjusting the parameters of the statistical model or using any other reaction model. These results might indicate the initial quantum mechanical delay (due to the slow non-exponential decay of the exit channel dinuclear states) followed by a fast exponential decay with a large Breit-Wigner width resulting in higher temperatures for the ensembles of the residual nuclei.

We thank R. Vandenbosch (University of Washington), J. N. De (Saha Institute of Nuclear Physics, Kolkata, India) and S. K. Samaddar (Saha Institute of Nuclear Physics, Kolkata, India) for a critical reading and useful discussion of our manuscript. We thank D. R. Chakrabarty (Nuclear Physics Division, Bhabha Atomic Research Center, Mumbai, India) for performing 'CASCADE' code calculations and useful discussions. A. De acknowledges the supports received from UGC (Ref No. F.PSW-011/07-08(ERO)) and collaborative research scheme No. UGC-DAE-CSR-KC/CRS/2009/NP-TIFR02.

**Figures**

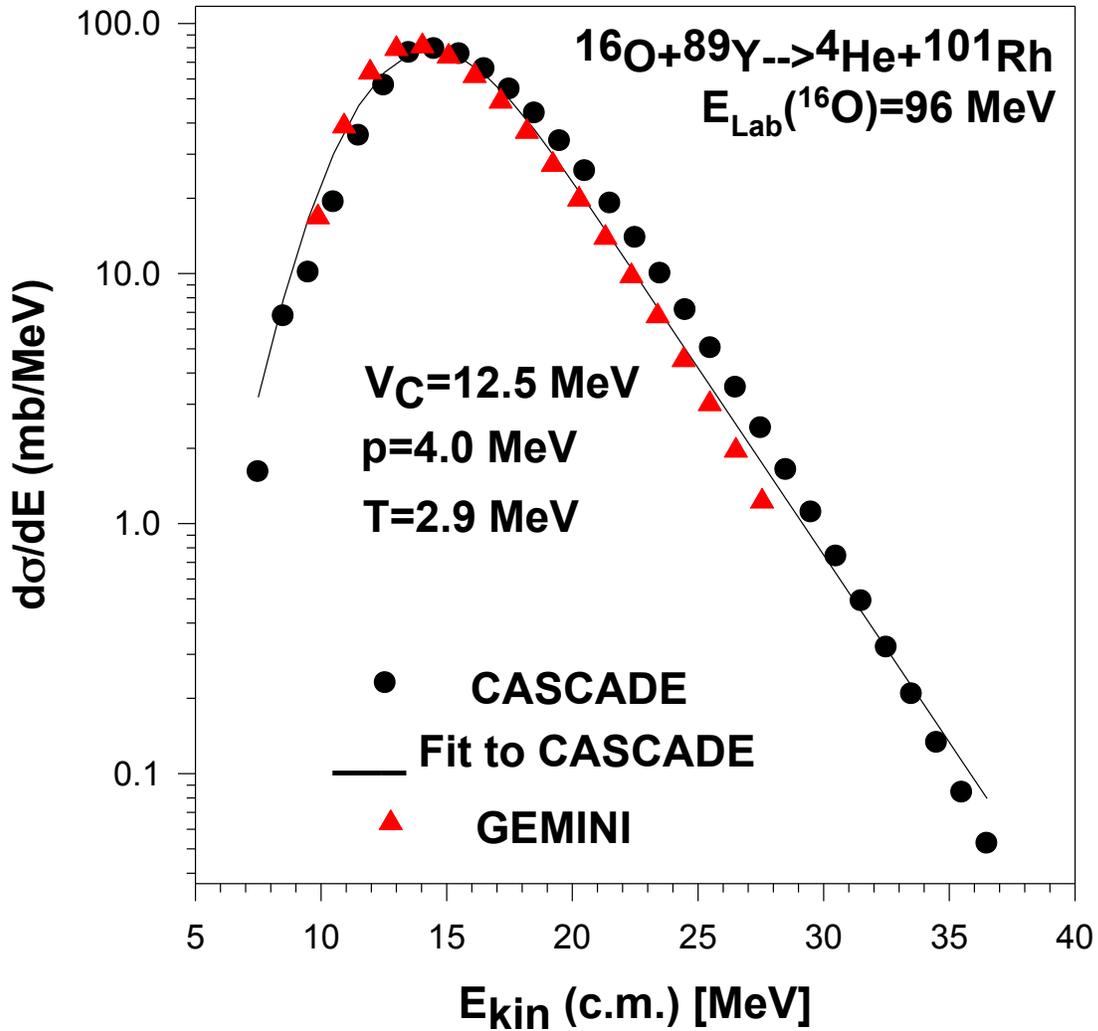

**Figure 1.** (Color online) Statistical model 'CASCADE' and 'GEMINI' code calculations of $^4$He spectrum from the $^{16}$O+$^{89}$Y reaction at $E_{Lab}(^{16}O)$ = 96 MeV and the corresponding fit (smooth line) using eq. (1). The 'GEMINI' code calculations have been normalized to match with the 'CASCADE' code calculations.

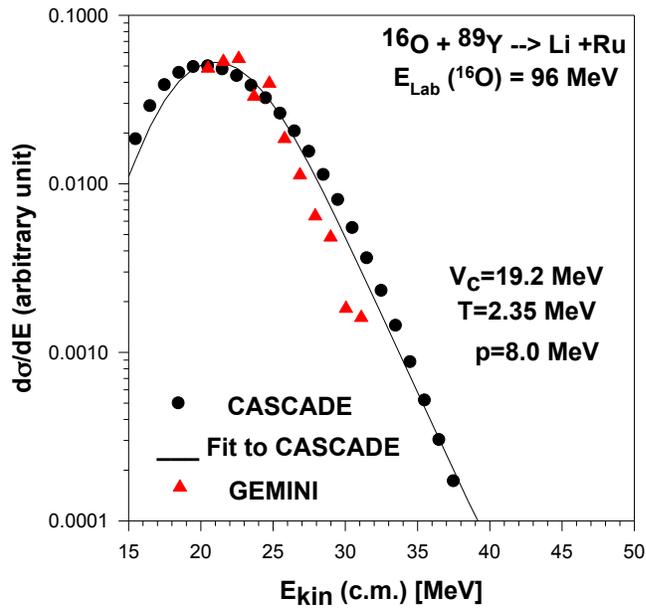

**Figure 2.** (Color online) Statistical model 'CASCADE' and 'GEMINI' code calculations of lithium spectrum from the $^{16}O+^{89}Y$ reaction at $E_{Lab}(^{16}O)$ = 96 MeV and the corresponding fit (smooth line) using eq. (1). The 'GEMINI' calculations have been normalized to match with the 'CASCADE' calculations.

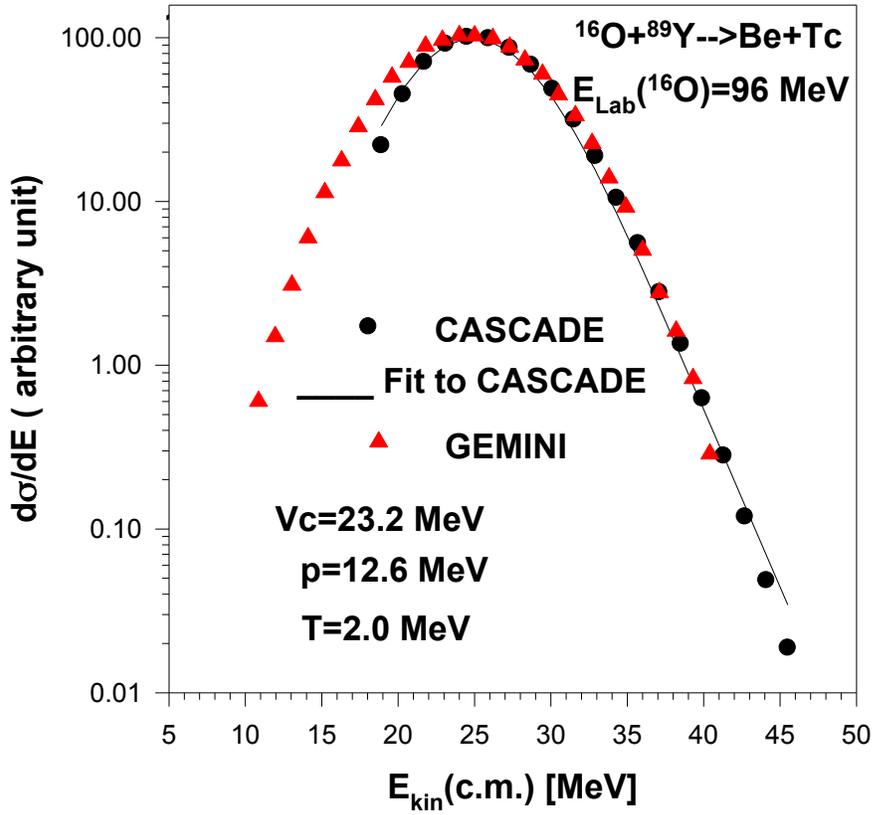

**Figure 3.** (Color online) Statistical model 'CASCADE' and 'GEMINI' code calculations of beryllium spectrum from the $^{16}O+^{89}Y$ reaction at $E_{Lab}(^{16}O)$ = 96 MeV and the corresponding fit (smooth line\) using eq. (1). The 'GEMINI' calculations have been normalized to match with the 'CASCADE' calculations.

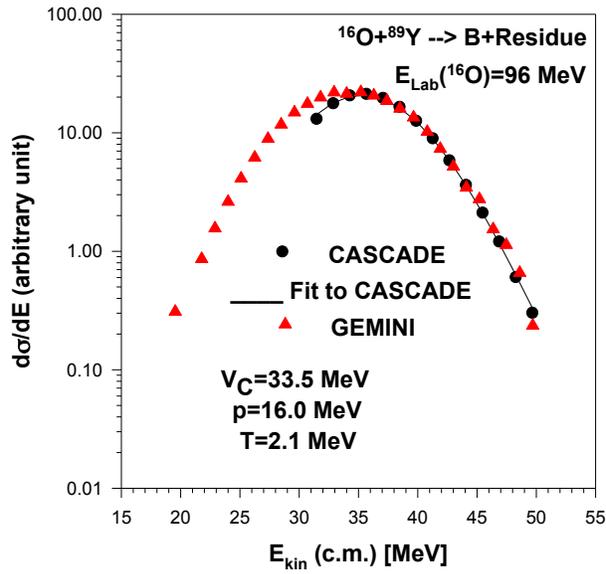

**Figure 4.** (Color online) Statistical model 'CASCADE' and 'GEMINI' code calculations of boron spectrum from the $^{16}O+^{89}Y$ reaction at $E_{Lab}(^{16}O) = 96$ MeV and the corresponding fit (smooth line) using eq. (1). The 'GEMINI' calculations have been normalized to match with the 'CASCADE' calculations.

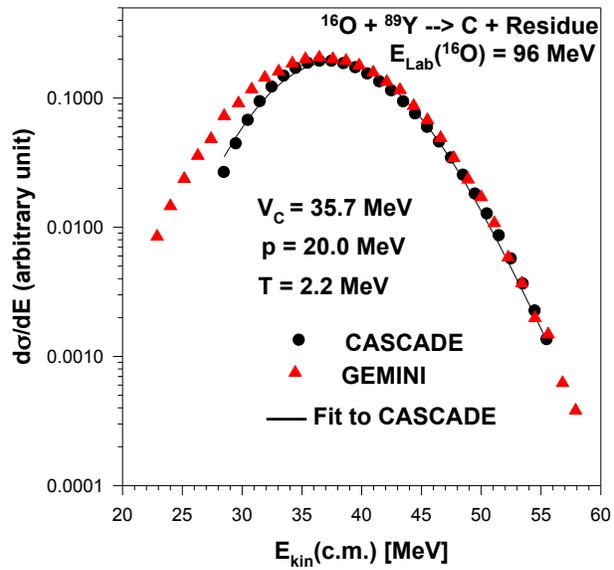

**Figure 5.** (Color online) Statistical model 'CASCADE' and 'GEMINI' code calculations of carbon spectrum from the $^{16}$O+$^{89}$Y reaction at $E_{Lab}(^{16}O)$ = 96 MeV and the corresponding fit (smooth line) using eq. (1). The 'GEMINI' calculations have been normalized to match with the 'CASCADE' calculations.

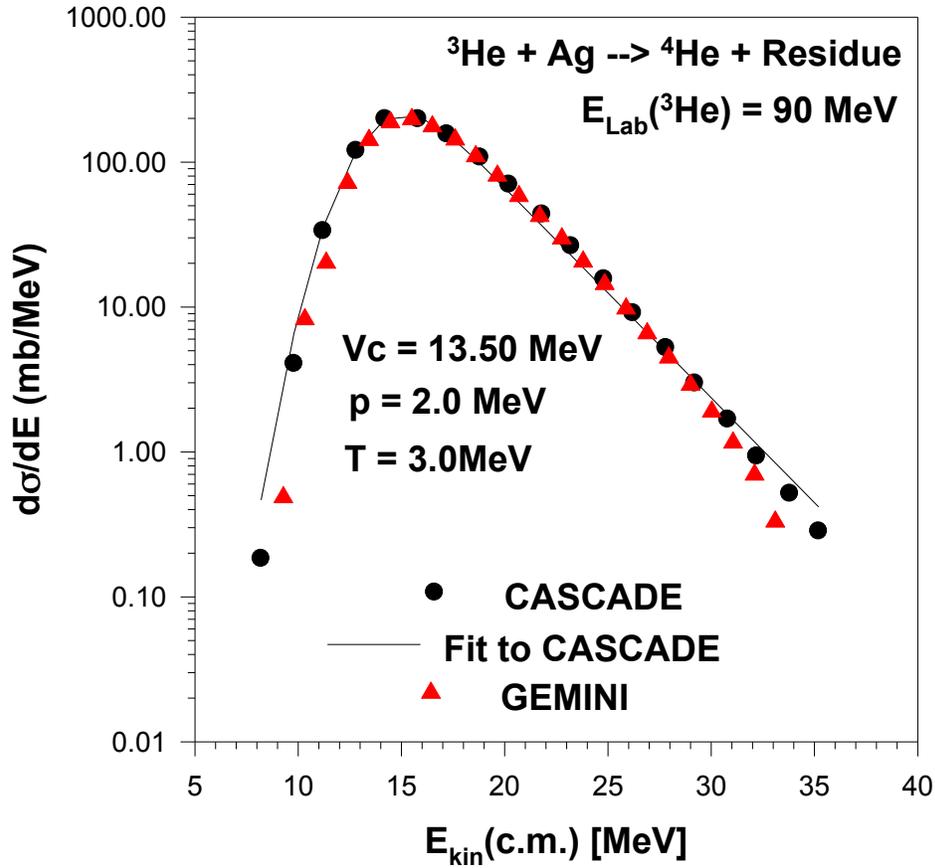

**Figure 6.** (Color online) Statistical model 'CASCADE' and 'GEMINI' code calculations of $^4$He spectrum from the $^3$He+Ag reaction at $E_{Lab}(^3He) = 90$ MeV and the corresponding fit (smooth line) using eq. (1). The 'GEMINI' calculations have been normalized to match with the 'CASCADE' calculations.

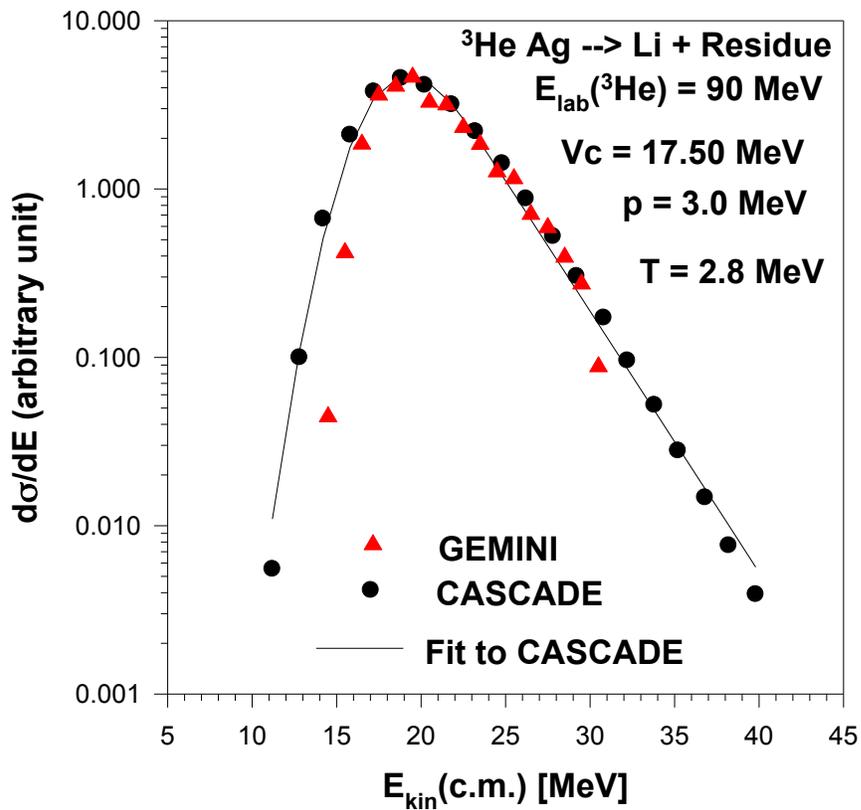

**Figure 7.** (Color online) Statistical model 'CASCADE' and 'GEMINI' code calculations of lithium spectrum from the $^3$He+Ag reaction at $E_{Lab}(^3He)$ = 90 MeV and the corresponding fit (smooth line) using eq. (1). The 'GEMINI' calculations have been normalized to match with the 'CASCADE' calculations.

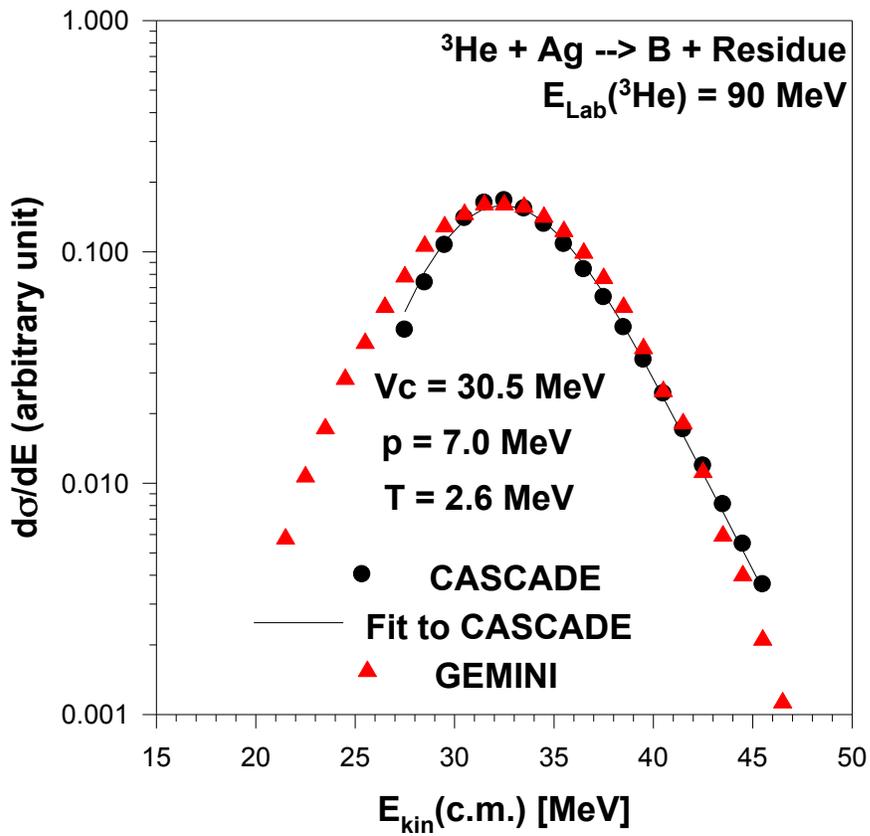

**Figure 8.** (Color online) Statistical model 'CASCADE' and 'GEMINI' code calculations of boron spectrum from the $^3$He+Ag reaction at $E_{Lab}(^3He)$ = 90 MeV and the corresponding fit (smooth line) using eq. (1). The 'GEMINI' calculations have been normalized to match with the 'CASCADE' calculations.

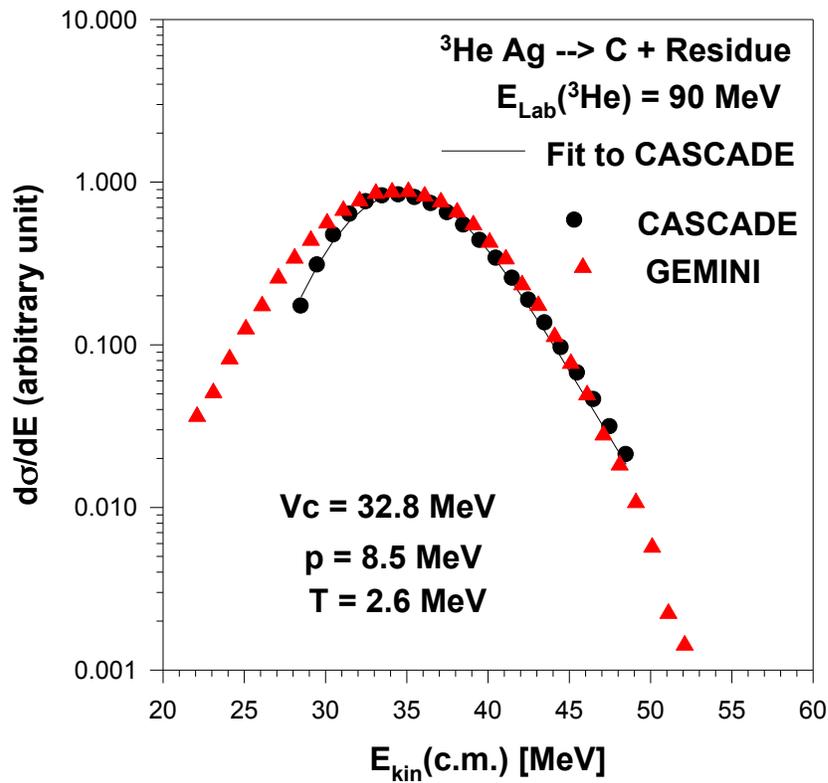

**Figure 9.** (Color online) Statistical model 'CASCADE' and 'GEMINI' code calculations of carbon spectrum from the $^3$He+Ag reaction at E($^3$He)$_{lab}$ = 90 MeV and the corresponding fit (smooth line) using eq. (1). The 'GEMINI' calculations have been normalized to match with the 'CASCADE' calculations.

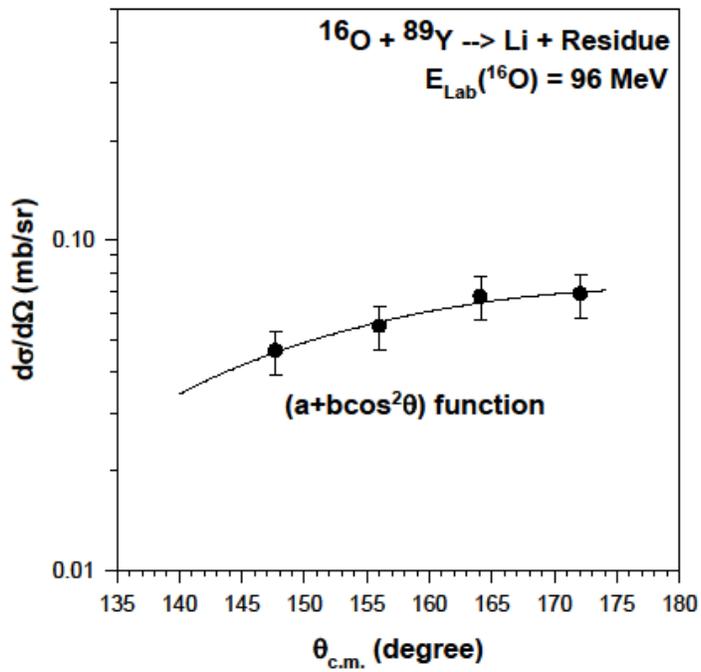

**Figure 10.** (Color online) Angular distribution of lithium particles integrated over the exit channel excitation energy region ($0 \leq E_X \leq 43$ MeV) in the center of mass frame. The solid line is $(a + b\cos^2\theta)$ function fit.

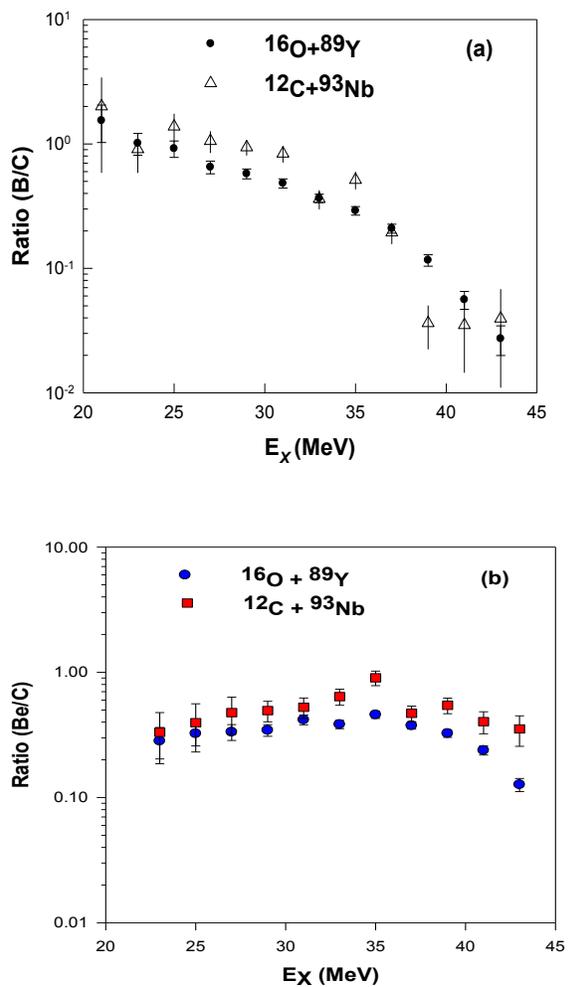

**Figure 11.** (Color online) Angle-integrated ratios of (a) B/C and (b) Be/C versus exit channel excitation energy for $^{16}O+^{89}Y$ and $^{12}C+^{93}Nb$ reactions producing $^{105}Ag$ at $E_X$=76 MeV.

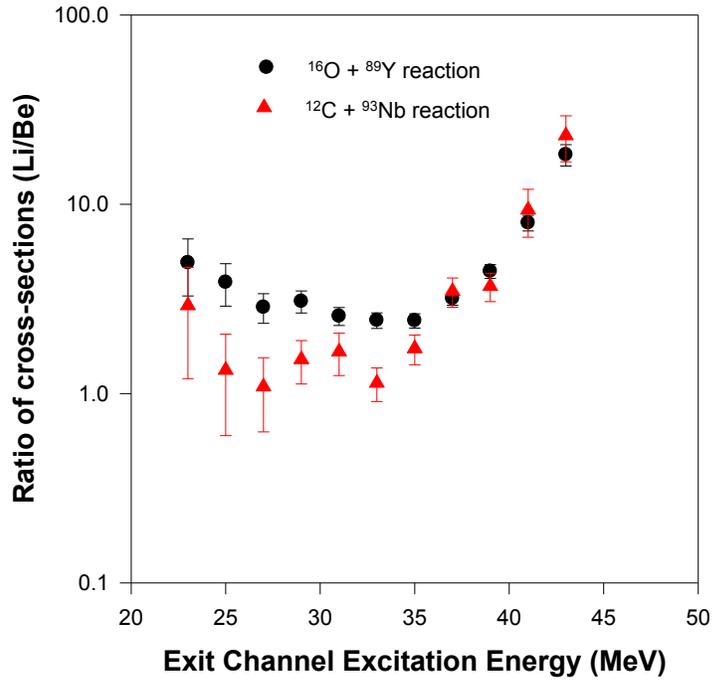

**Figure 12.** (Color online) Angle-integrated ratios of Li/Be versus exit channel excitation energy for $^{16}$O+$^{89}$Y and $^{12}$C+$^{93}$Nb reactions producing $^{105}$Ag at $E_X$=76 MeV.

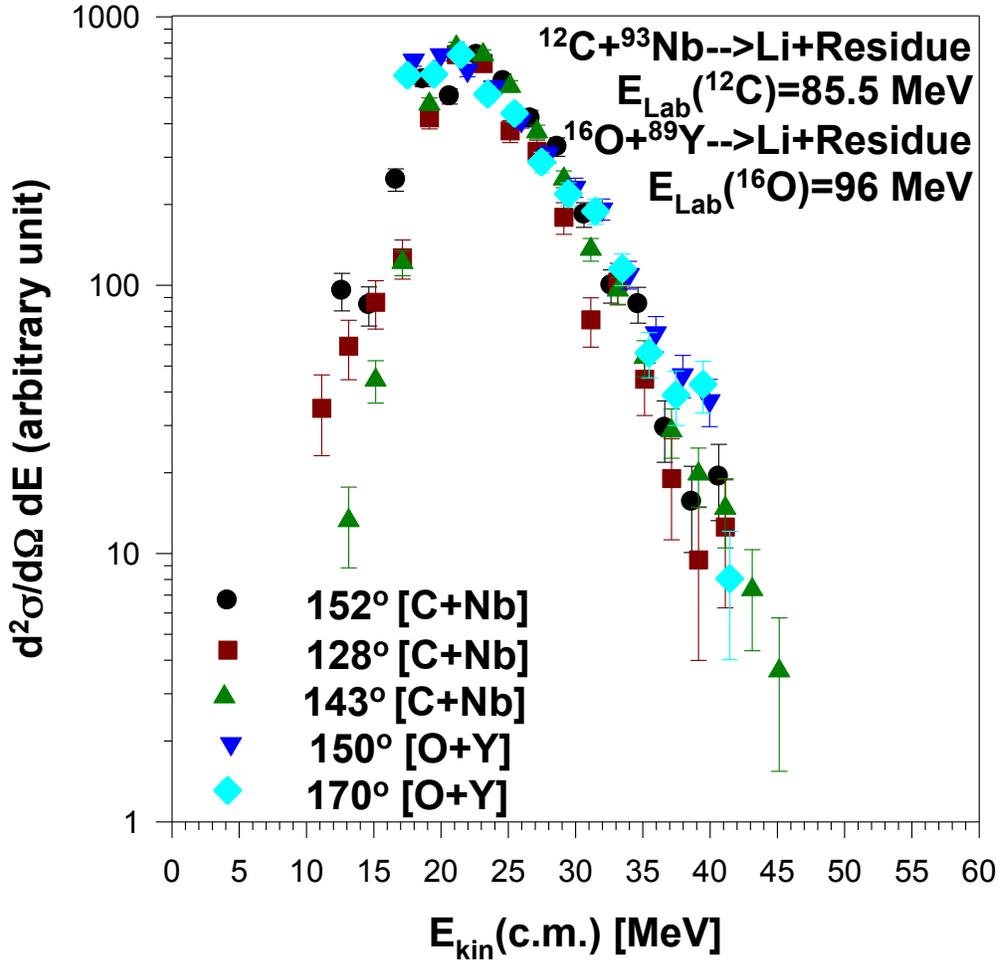

**Figure 13.** (Color online) Overlay plots of lithium spectra (after suitable normalizations) at different angles for $^{12}C+^{93}Nb$ reaction and $^{16}O+^{89}Y$ reaction forming the $^{105}Ag$ at $E_X=76$ MeV.

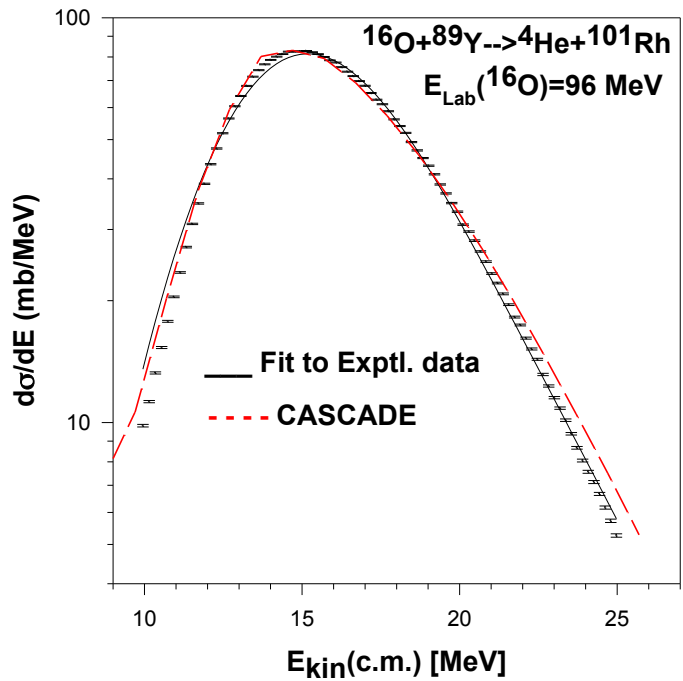

**Figure 14.** (Color online) Overlay plots of experimental and theoretical statistical model spectra for alpha particles from the $^{16}$O+$^{89}$Y reaction at $E_{Lab}(^{16}O)$ = 96 MeV and the corresponding fit to the experimental spectrum using eq. (1). The theoretical spectrum has been normalised and shifted with respect to the corresponding experimental spectrum to overlay their peak positions. Solid black curve represents fit to the experimental data points. Dashed red curve shows statistical model CASCADE code calculations.

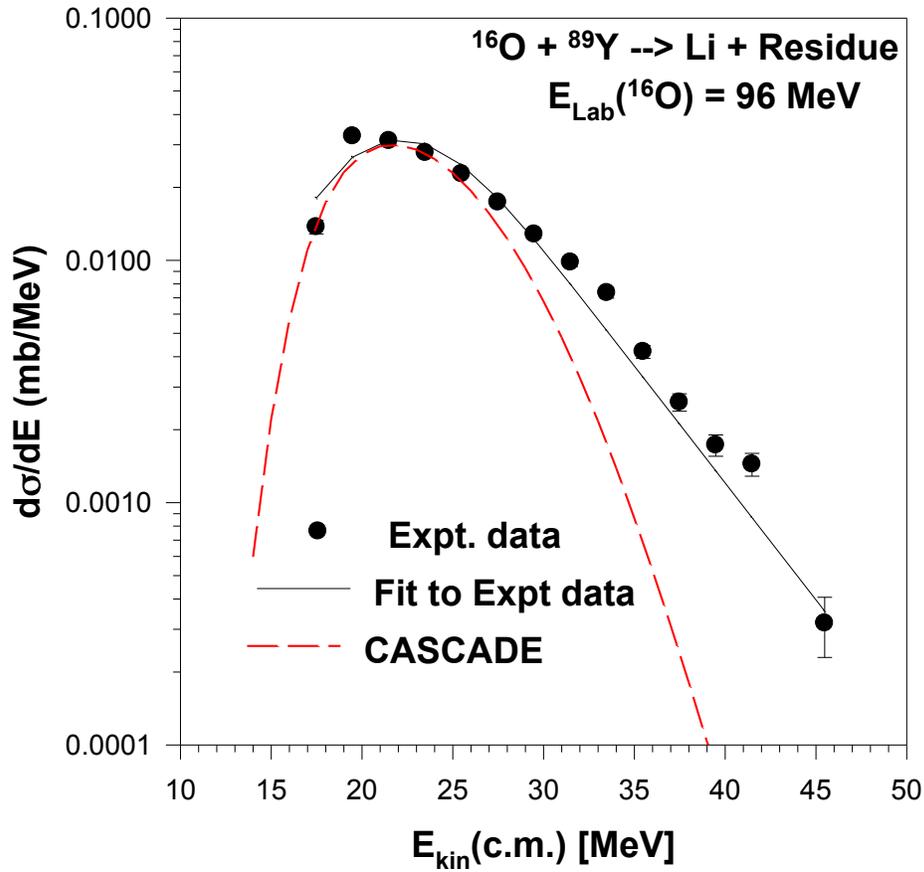

**Figure 15.** (Color online) Overlay plots of experimental and theoretical statistical model spectra for lithium particles from the $^{16}O+^{89}Y$ reaction at $E_{Lab}(^{16}O) = 96$ MeV and the corresponding fit to the experimental spectrum using eq. (1). The theoretical spectrum has been normalised and shifted with respect to the corresponding experimental spectrum to overlay their peak positions. Solid black curve represents fit to the experimental data points. Dashed red curve shows statistical model CASCADE code calculations.

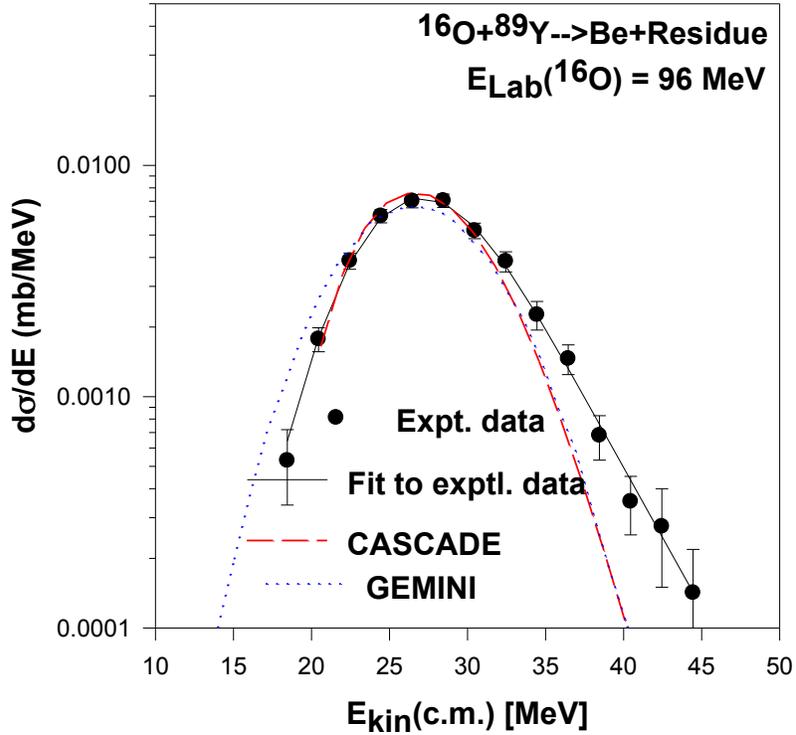

**Figure 16.** (Color online) Overlay plots of experimental and theoretical statistical model spectra for beryllium particles from the $^{16}O+^{89}Y$ reaction at $E_{Lab}(^{16}O)$ = 96 MeV and the corresponding fit to the experimental spectrum using eq. (1). The theoretical spectrum has been normalised and shifted with respect to the corresponding experimental spectrum to overlay their peak positions. Solid black curve represents fit to the experimental data points. Dashed red curve and dotted blue curve show statistical model CASCADE code and GEMINI code calculations respectively.

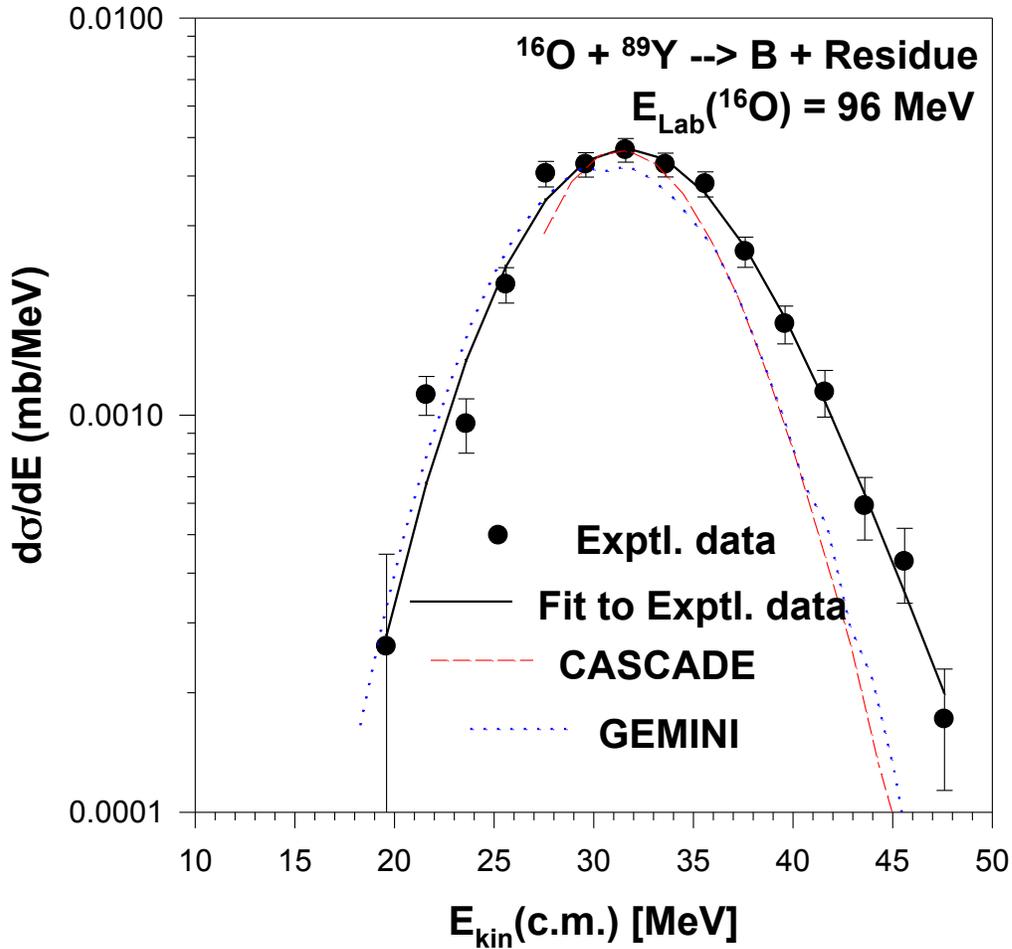

**Figure 17.** (Color online) Overlay plots of experimental and theoretical statistical model spectra for boron particles from the $^{16}$O+$^{89}$Y reaction at $E_{Lab}(^{16}O) = 96$ MeV and the corresponding fit to the experimental spectrum using eq. (1). The theoretical spectrum has been normalised and shifted with respect to the corresponding experimental spectrum to overlay their peak positions. Solid black curve represents fit to the experimental data points. Dashed red curve and dotted blue curve show statistical model CASCADE code and GEMINI code calculations respectively.

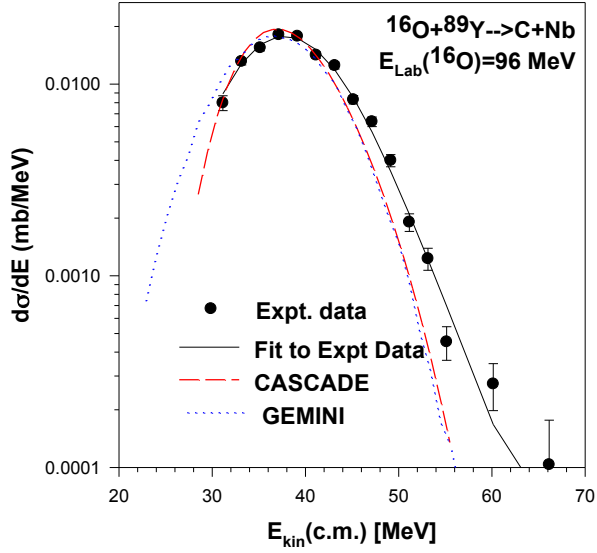

**Figure 18.** (Color online) Overlay plots of experimental and theoretical statistical model spectra for carbon particles from the $^{16}$O+$^{89}$Y reaction at $E_{Lab}(^{16}O)$ = 96 MeV and the corresponding fit to the experimental spectrum using eq. (1). The theoretical spectrum has been normalised and shifted with respect to the corresponding experimental spectrum to overlay their peak positions. Solid black curve represents fit to the experimental data points. Dashed red curve and dotted blue curve show statistical model CASCADE code and GEMINI code calculations respectively.

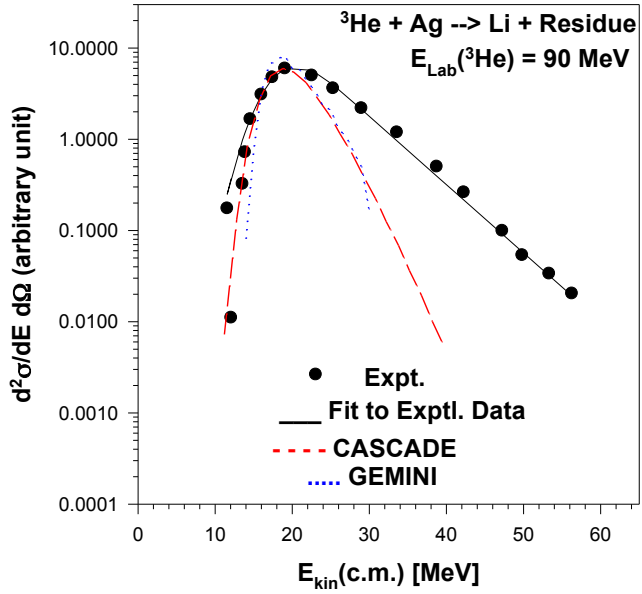

**Figure 19.** (Color online) Overlay plots of experimental and theoretical statistical model spectra for lithium particles from the $^3$He+Ag reaction at $E_{Lab}(^3He)$ = 90 MeV and the corresponding fit to the experimental spectrum using eq. (1). The theoretical spectrum has been normalised and shifted with respect to the corresponding experimental spectrum to overlay their peak positions. Solid black curve represents fit to the experimental data points. Dashed red curve and dotted blue curve show statistical model CASCADE code and GEMINI code calculations respectively.

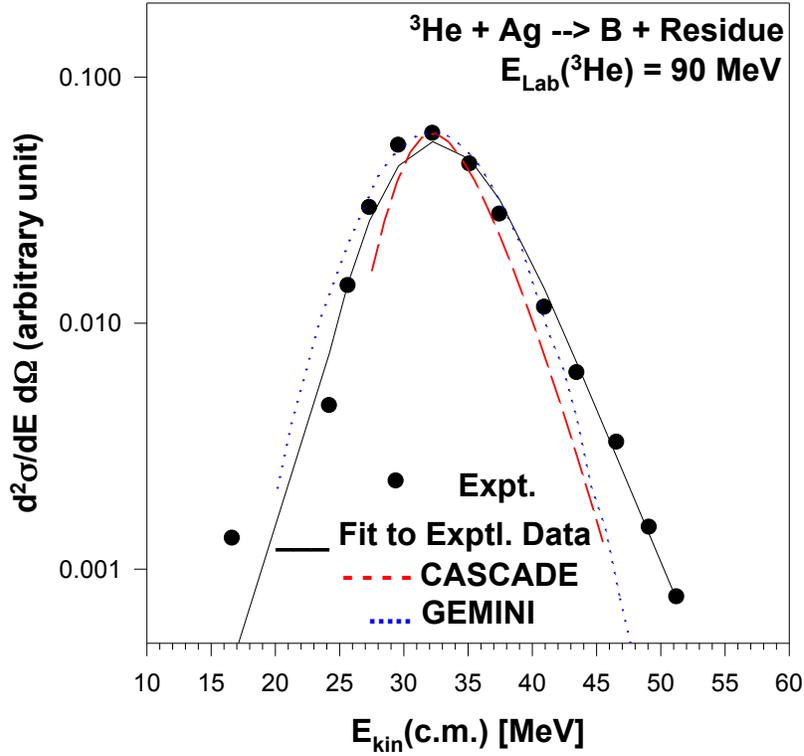

**Figure 20.** (Color online) Overlay plots of experimental and theoretical statistical model spectra for boron particles from the $^3$He+Ag reaction at $E_{Lab}(^3He)$ = 90 MeV and the corresponding fit to the experimental spectrum using eq. (1). The theoretical spectrum has been normalised and shifted with respect to the corresponding experimental spectrum to overlay their peak positions. Solid black curve represents fit to the experimental data points. Dashed red curve and dotted blue curve show statistical model CASCADE code and GEMINI code calculations respectively.

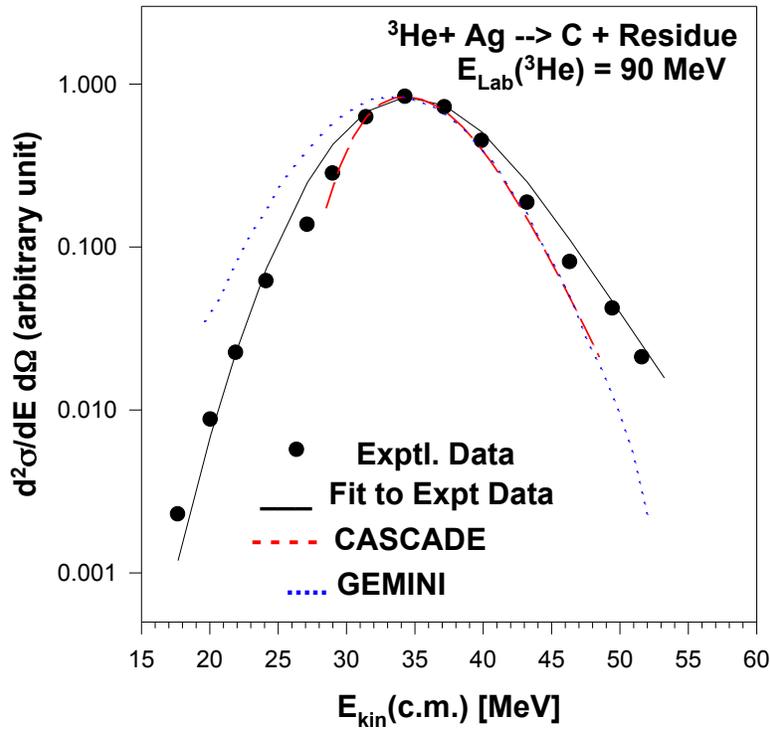

**Figure 21.** (Color online) Overlay plots of experimental and theoretical statistical model spectra for carbon particles from the $^3$He+Ag reaction at $E_{Lab}(^3He)$ = 90 MeV and the corresponding fit to the experimental spectrum using eq. (1). The theoretical spectrum has been normalised and shifted with respect to the corresponding experimental spectrum to overlay their peak positions. Solid black curve represents fit to the experimental data points. Dashed red curve and dotted blue curve show statistical model CASCADE code and GEMINI code calculations respectively.

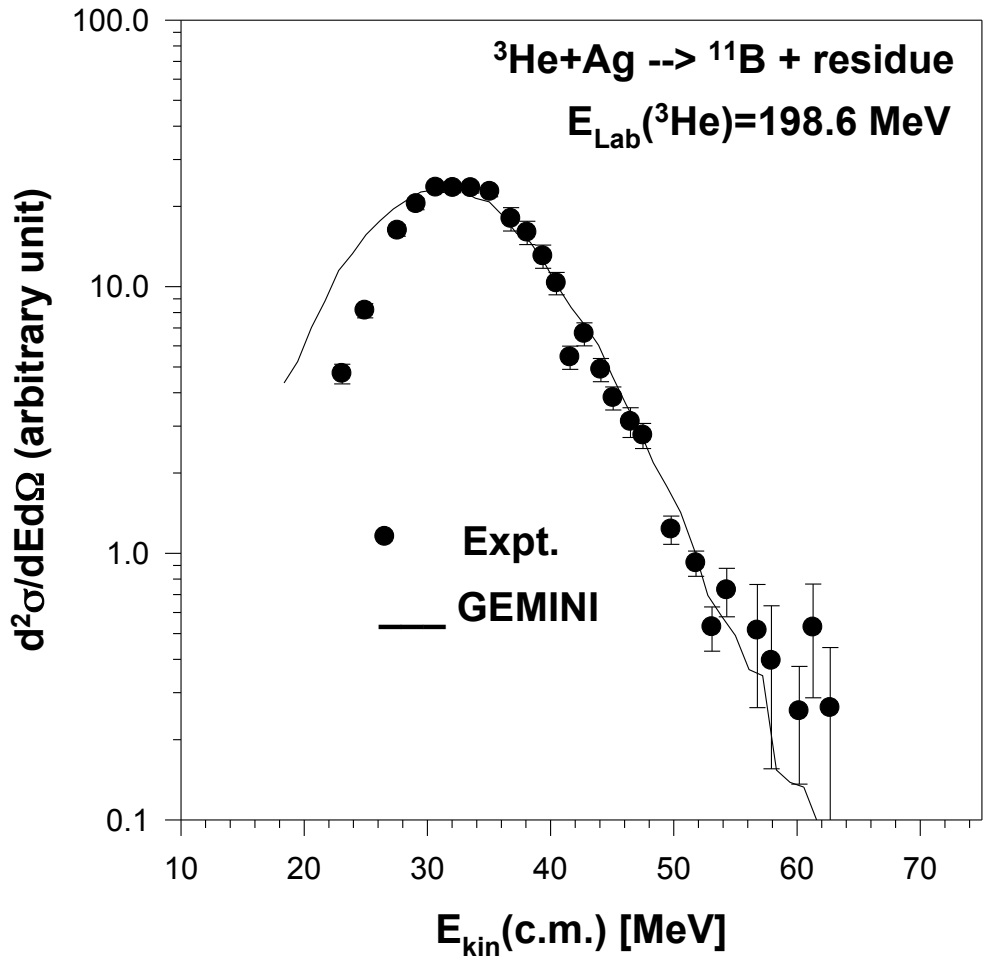

**Figure 22.** Overlay plots of experimental and theoretical statistical model spectra for boron particles from the $^3$He+Ag reaction at $E_{Lab}(^3He)$ = 198.6 MeV. The theoretical (GEMINI) spectrum has been normalised and shifted with respect to the corresponding experimental spectrum to overlay their peak positions.

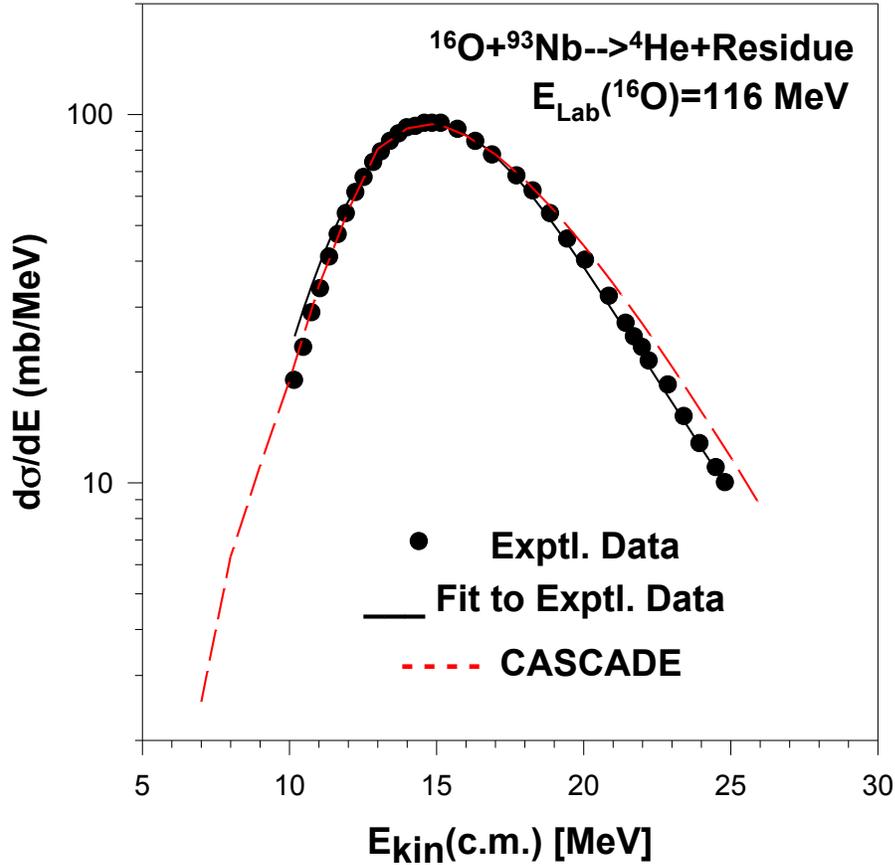

**Figure 23.** (Color online) Overlay plots of experimental and theoretical statistical model spectra for alpha particles from the $^{16}$O+$^{93}$Nb reaction at $E_{Lab}(^{16}O)$ = 116 MeV and the corresponding fit to the experimental spectrum using eq. (1). The theoretical spectrum has been normalised and shifted with respect to the corresponding experimental spectrum to overlay their peak positions. Solid black curve represents fit to the experimental data points. Dashed red curve shows statistical model CASCADE code calculations.

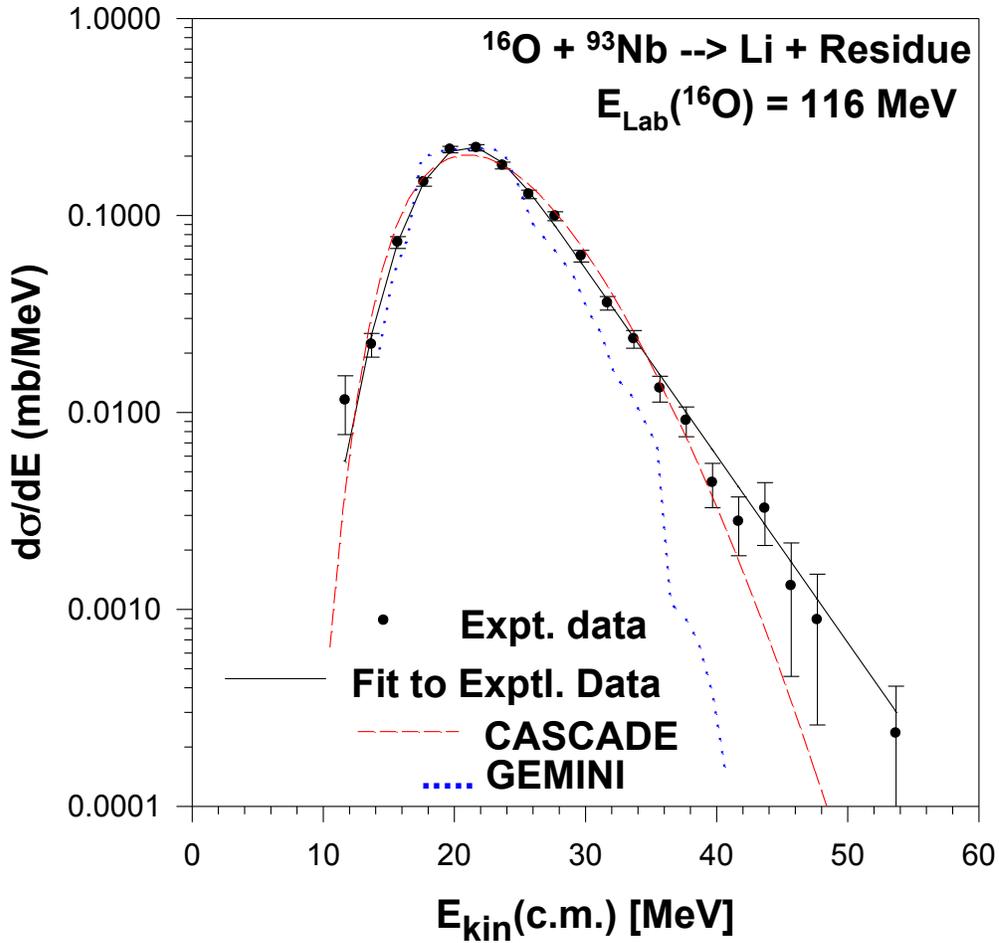

**Figure 24.** (Color online) Overlay plots of experimental and theoretical statistical model spectra for lithium particles from the $^{16}$O+$^{93}$Nb reaction at $E_{Lab}(^{16}O)$ = 116 MeV and the corresponding fit to the experimental spectrum using eq. (1). The theoretical spectra ('CASCADE' and 'GEMINI') have been normalised and shifted with respect to the corresponding experimental spectrum to overlay their peak positions. Solid black curve represents fit to the experimental data points. Dashed red curve and dotted blue curve show statistical model CASCADE code and GEMINI code calculations respectively.

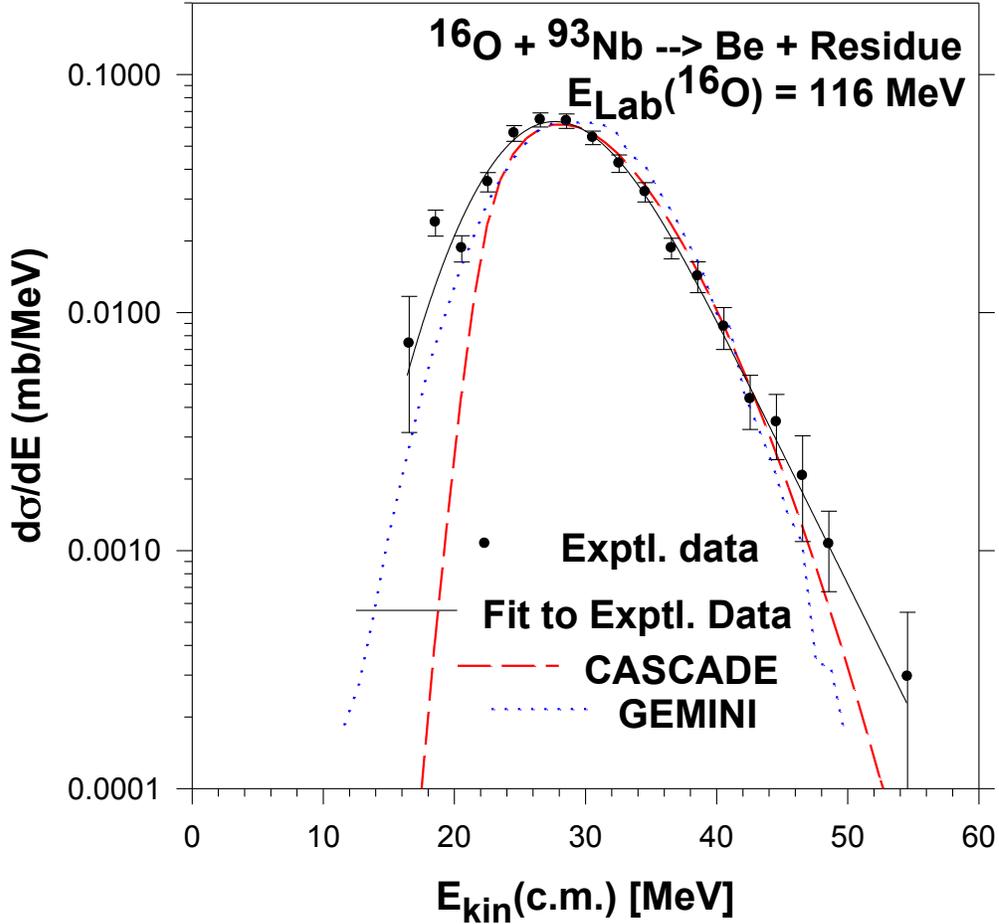

**Figure 25.** (Color online) Overlay plots of experimental and theoretical statistical model spectra for beryllium particles from the $^{16}O+^{93}Nb$ reaction at $E_{Lab}(^{16}O)$ = 116 MeV and the corresponding fit to the experimental spectrum using eq. (1). The theoretical spectra ('CASCADE' and 'GEMINI') have been normalised and shifted with respect to the corresponding experimental spectrum to overlay their peak positions. Solid black curve represents fit to the experimental data points. Dashed red curve and dotted blue curve show statistical model CASCADE code and GEMINI code calculations respectively.

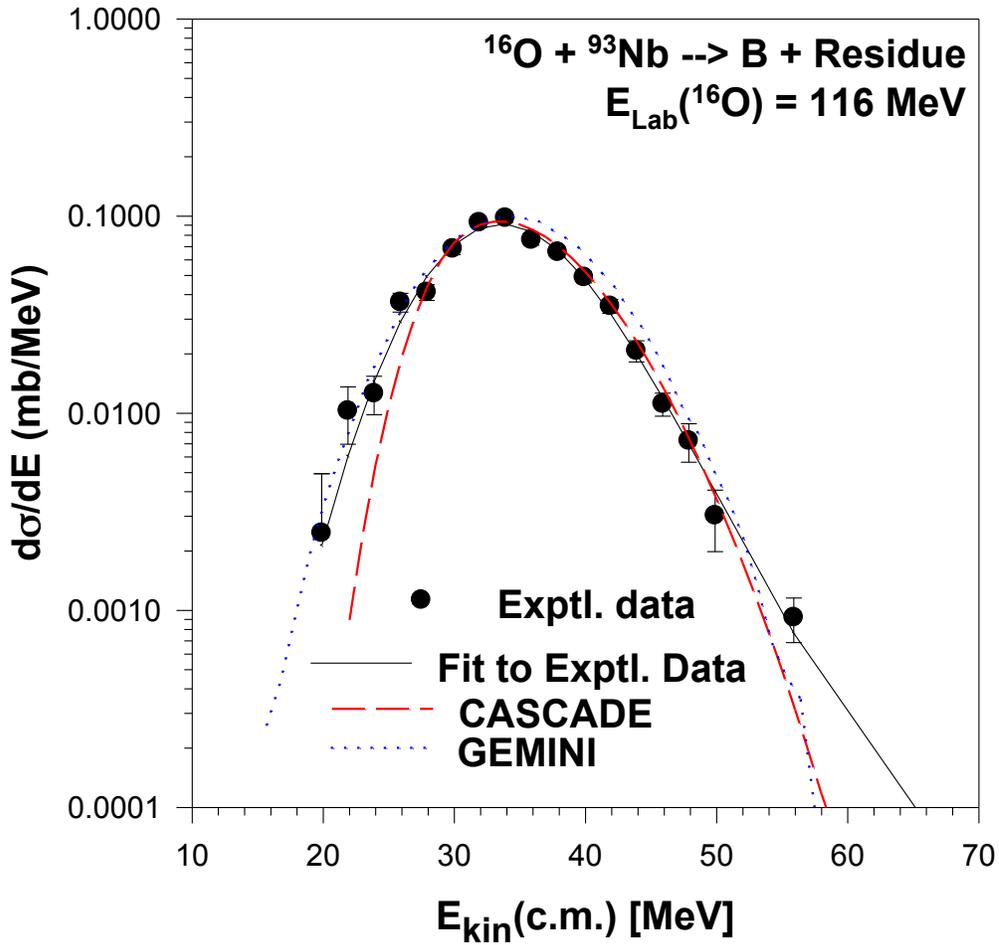

**Figure 26.** (Color online) Overlay plots of experimental and theoretical statistical model spectra for boron particles from the $^{16}$O+$^{93}$Nb reaction at $E_{Lab}(^{16}O) = 116$ MeV and the corresponding fit to the experimental spectrum using eq. (1). The theoretical spectra ('CASCADE' and 'GEMINI') have been normalised and shifted with respect to the corresponding experimental spectrum to overlay their peak positions. Solid black curve represents fit to the experimental data points. Dashed red curve and dotted blue curve show statistical model CASCADE code and GEMINI code calculations respectively.

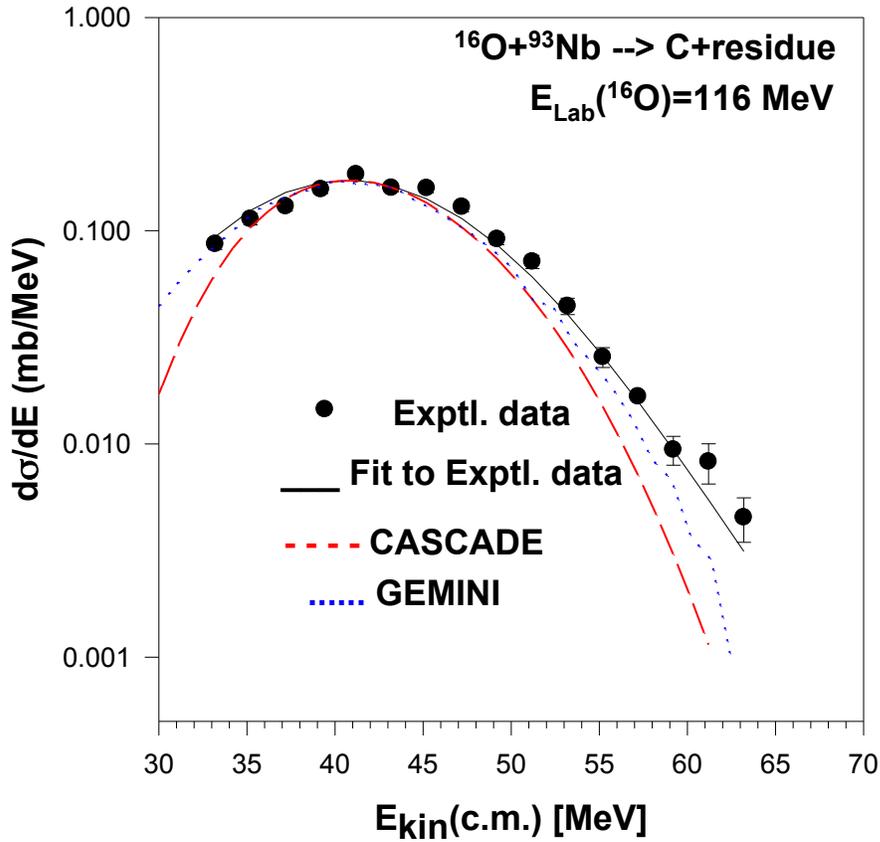

**Figure 27.** (Color online) Overlay plots of experimental and theoretical statistical model spectra for carbon particles from the $^{16}$O+$^{93}$Nb reaction at $E_{Lab}(^{16}O)$ = 116 MeV and the corresponding fit to the experimental spectrum using eq. (1). The theoretical spectra ('CASCADE' and 'GEMINI') have been normalised and shifted with respect to the corresponding experimental spectrum to overlay their peak positions. Solid black curve represents fit to the experimental data points. Dashed red curve and dotted blue curve show statistical model CASCADE code and GEMINI code calculations respectively.

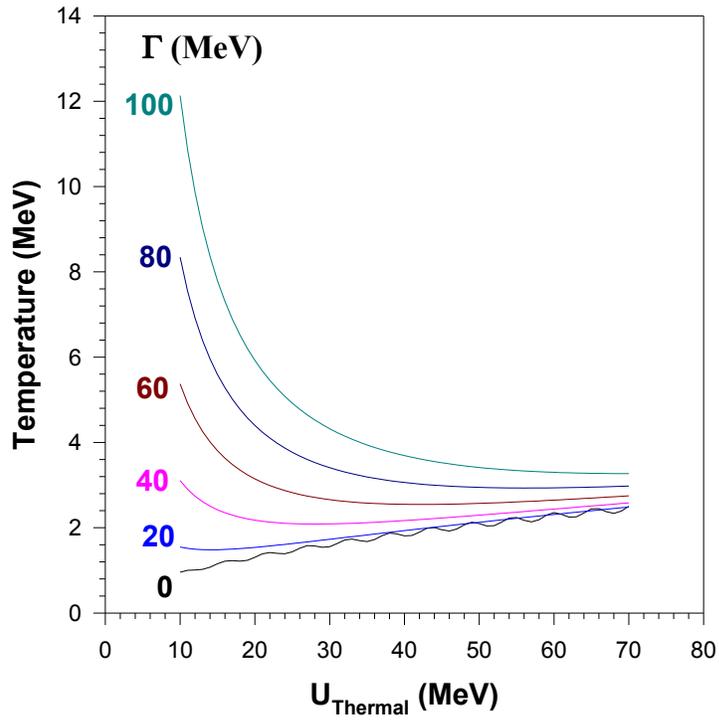

**Fig. 28.** (Color online) Plot of temperature (T) versus thermal energy (U) for different values of Breit-Wigner width (Γ) using eq. (3).

Table 1. Comparison of temperature (T) and p-parameters derived from the experimental and statistical model spectra.

| System studied | Projectile Energy and $E_X$ of CN (MeV) | Fragment Investi-gated | Derived parameters from | | | |
|---|---|---|---|---|---|---|
| | | | Experimental data | | Statistical model calculation | |
| | | | P (MeV) | T (MeV) | P (MeV) | T (MeV) |
| $^{16}$O + $^{89}$Y | 96 ($^{16}$O) $^{105}$Ag $E_X$=76 MeV | $^4$He | 4.0±0.4 | 2.90±0.15 | 4.0 | 2.9 |
| | | Li | 6.0±0.6 | 4.50±0.3 | 8.0 | 2.35 |
| | | Be | 6.8 ±1.0 | 3.6± 0.3 | 12.6 | 2.0 |
| | | B | 13.1±1.0 | 3.35±0.2 | 16.0 | 2.1 |
| | | C | 15.0±1.2 | 3.5±0.3 | 18.0 | 2.2 |
| $^{16}$O + $^{93}$Nb | 116 ($^{16}$O) $^{109}$In $E_X$=93.5 MeV | $^4$He | 3.5± 0.4 | 3.5±0.1 | 4.0 | 3.4 |
| | | Li | 3.9± 0.4 | 4.6±0.2 | 11.0 | 2.8 |
| | | Be | 10.9±1.0 | 3.9±0.2 | 19.0 | 2.4 |
| | | B | 11.8±1.2 | 3.6±0.2 | 25.0 | 2.3 |
| | | C | 23.0±1.0 | 3.5±0.2 | 27.0 | 2.0 |
| $^3$He + Ag | 90 ($^3$He) $E_X$~82 MeV with broad distribution. | $^4$He | 2.0±0.2 | 3.0±0.15 | 2.0 | 3.0 |
| | | Li | 3.0±0.3 | 5.8±0.3 | 3.0 | 2.8 |
| | | B | 8.0±1.0 | 3.5±0.2 | 7.0 | 2.6 |
| | | C | 10.9±1.0 | 3.5±0.3 | 8.5 | 2.6 |